\documentclass[11pt]{article}%
\usepackage{amsmath}
\usepackage{amssymb}
\usepackage{amsfonts}

\usepackage{cite}
\usepackage{graphicx}
\usepackage{float}
\usepackage{longtable}
\usepackage[utf8]{inputenc}
\usepackage{hyperref}

\usepackage{algorithmicx}
\usepackage[ruled,vlined]{algorithm2e}
\SetAlgoCaptionSeparator{ }

\usepackage{fullpage}

\usepackage{array}
\newcolumntype{C}[1]{>{\centering\arraybackslash}m{#1}}

\newcommand{\fref}[1]{Fig.~\ref{#1}}

\newcommand{\tref}[1]{Table~\ref{#1}}
\newcommand{\sref}[1]{Section~\ref{#1}}

\providecommand{\U}[1]{\protect\rule{.1in}{.1in}}
\newtheorem{theorem}{Theorem}

\newenvironment{proof}[1][Proof]{\textbf{#1.} }{\ \rule{0.5em}{0.5em}}

\begin{document}

\title{Theory of Quantum Gravity Information Processing}
\author{Laszlo Gyongyosi\thanks{School of Electronics and Computer Science, University of Southampton, Southampton SO17 1BJ, U.K., and Department of Networked Systems and Services, Budapest University of Technology and Economics, 1117 Budapest, Hungary, and MTA-BME Information Systems Research Group, Hungarian Academy of Sciences, 1051 Budapest, Hungary.}
\and Sandor Imre\thanks{Department of Networked Systems and Services, Budapest University of Technology and Economics, 1117 Budapest, Hungary.}}

\date{}

\maketitle
\begin{abstract}
The theory of quantum gravity is aimed to fuse general relativity with quantum theory into a more fundamental framework. The space of quantum gravity provides both the non-fixed causality of general relativity and the quantum uncertainty of quantum mechanics. In a quantum gravity scenario, the causal structure is indefinite and the processes are causally non-separable. Here, we provide a model for the information processing structure of quantum gravity. We show that the quantum gravity environment is an information resource-pool from which valuable information can be extracted. We analyze the structure of the quantum gravity space and the entanglement of the space-time geometry. We study the information transfer capabilities of quantum gravity space and define the quantum gravity channel. We reveal that the quantum gravity space acts as a background noise on the local environment states. We characterize the properties of the noise of the quantum gravity space and show that it allows the separate local parties to simulate remote outputs from the local environment state, through the process of remote simulation.
\end{abstract}

\section{Introduction}
\label{sec1}
In general relativity, processes and events are causally non-separable because the causal structure of space-time geometry is non-fixed. In a non-fixed causality structure, the sequence of time steps has no interpretable meaning. In our macroscopic world, events and processes are distinguishable in time and, thus, causally separable because the space-time geometry has a deterministic causality structure. The meaning of time evolution is also non-vanishing and has an interpretable notion in the microscopic world of quantum mechanics \cite{ref1,ref2,ref3,ref4,ref5,ref6,ref7}. It is precisely the reason why classical and quantum computations are evolved by a sequence of time steps and why the term time has an interpretable and plausible meaning in the macro- and microscopic levels. A fundamental difference between the nature of events of general relativity and quantum mechanics is that although the theory of general relativity provides a non-fixed causal space-time structure with deterministic events, in quantum mechanics, the space-time geometry has a fixed, deterministic causality structure whereas the events are nondeterministic. Quantum gravity is provided to fill the gap between these two fundamentally different theories. The theory of quantum gravity combines the results of general relativity with quantum mechanics to construct a more general framework. In a model of quantum gravity, the causal structure is non-fixed, and the events are probabilistic. In this quantum gravity environment, the computations and the information processing steps are interpreted without the notion of time evolution. This space-time structure allows us to perform quantum gravity computations and to build quantum gravity computers, which fuse the extreme power of quantum computations and the non-fixed causality structure of general relativity \cite{ref4}. The space of quantum gravity can be further exploited in quantum communication protocols, in quantum AI, in quantum error correction, and particularly in the development of quantum computers \cite{ref8,ref9,ref10,ref11,ref12,ref13,ref14,ref15,ref16,ref17,ref18,ref19,ref20,ref21,ref22,ref23,ref24,ref25,ref26,ref27,ref27b,ref29,ref30,ref31,ref32,ref33,ref41,ref42,ref43,ref44,ref45,ref46,ref47,ref48,ref49,ref50,ref51,ref52,ref53,ref54,ref55, pres,har, aar, far1,far2,qcomputer, refibm,nagyp}.

Some related works are as follows. For a theoretical background on quantum gravity computations, we suggest \cite{ref2,ref3,ref4,ref5,ref6,ref7,ref8,ref12}. For a work on experimental superposition of orders of quantum gates, see \cite{add1}. The problem of exponential communication complexity advantage from quantum superposition of the direction of communication has been studied in \cite{add2}. In \cite{add3}, the authors proposed an experimental verification of an indefinite causal order. The works of \cite{add1,add2,add3} also provide the basic indefinite causal structure and reveal its experimental aspects. On the problem of the general quantum interference principle and duality computer, see \cite{add4}. This work also defines a quantum computation model that can be used to simulate an indefinite causal structure. For the structural attributes of an indefinite causal structure, see also \cite{add5}. A special computational framework has been defined in \cite{add6}. This work also studies the problem of quantifiable simulation of quantum computation beyond stochastic ensemble computation. In \cite{add7}, the authors studied the emulation of quantum algorithms at high-precision and high-throughput, and also evaluated a computational model.

Besides the attractive properties of quantum gravity theory, the appropriate characterization of the information processing structure of the quantum gravity space is still missing. Here, our aim was to provide a model for the information processing structure of quantum gravity. We show that the quantum gravity space acts as an information resource-pool and reveal that the entangled structure of the quantum gravity space stimulates a noisy map on the local environment states of independent, physically separated local maps. This background noise (a noisy map) of the quantum gravity space allows the local parties to simulate remote, physically separated processes in the quantum gravity space, in a probabilistic way. We call this process remote simulation, an event that can be accomplished only as a coin tossing in a fixed causality structure. We also study the entangled space-time structure of quantum gravity and define the partitions over which the information flow between the separated processes is possible. We characterize the properties of the quantum gravity channel and the information transmission capability of the quantum gravity space by the tools of quantum Shannon theory \cite{o1,o2,o6,o7,o9,o9b,o10,o11,o13,o14,o15,o16,o17,o18, ref34,ref35,ref36,ref37,ref38,ref39,ref40}. We introduce the terms quantum gravity memory and stimulated storage, which allow for the generation and storage of qubit entanglement exploiting the information resource-pool property of the quantum gravity space.

The novel contributions of our paper are as follows.

\begin{enumerate}
\item \textit{We provide a model for the information processing structure of quantum gravity.} 

\item \textit{We analyze the structure of the quantum gravity space and the entanglement of the space-time geometry. We study the information transfer capabilities of quantum gravity space and define the quantum gravity channel.} 

\item \textit{We reveal that the quantum gravity space acts as a background noise on the local environment states.} 

\item \textit{We characterize the properties of the noise of the quantum gravity space and show that it allows the separate local parties to simulate remote outputs from the local environment state, through the process of remote simulation.} 

\item \textit{We characterize the information transfer of the gravity space and the correlation measure functions of the gravity channel.}
\end{enumerate}

This paper is organized as follows. \sref{sec2} provides the entanglement structure of the quantum gravity space, the information resource-pool property of quantum gravity, and the structure of the quantum gravity channel. \sref{sec3} studies the information flow through the quantum gravity environment and characterizes the correlation measures. \sref{sec4} provides a quantum gravity memory and introduces the term stimulated storage. Finally, \sref{sec5} concludes the paper. Supplemental information is included in the Appendix.

\section{Structure of Information Processing}
\label{sec2}
\begin{theorem}
(Entangled structure of the quantum gravity environment). The space-time geometry (quantum gravity environment ${\rm {\mathcal G}}_{E} $) can formulate an entangled structure with $E_{i} B_{j} $, where $E_{i} $ is the local environment, and $B_{j} $ is the remote output of local maps ${\rm {\mathcal M}}_{1} $ and ${\rm {\mathcal M}}_{2} $, $i\ne j$. The $\rho _{{\rm {\mathcal G}}_{E} E_{i} B_{j} } $ entangled structure stimulates a non-fixed causality between the local processes ${\rm {\mathcal M}}_{1} $ and ${\rm {\mathcal M}}_{2} $.
\end{theorem}
\begin{proof}
The proofs throughout this work assume two qubit maps ${\rm {\mathcal M}}_{1} $ and ${\rm {\mathcal M}}_{2} $, with qubit quantum gravity environment state ${\rm {\mathcal G}}_{E} $. Specifically, the utilization of qubit channels is a required condition of the existence of a non-fixed causality structure between independent local completely positive, trace preserving (CPTP) maps ${\rm {\mathcal M}}_{1} $ and ${\rm {\mathcal M}}_{2} $, which follow from the property of the shift-and-multiply unitaries \cite{ref11}.

The local CPTP maps ${\rm {\mathcal M}}_{1} $ and ${\rm {\mathcal M}}_{2} $ are independent, physically separated maps, with uncorrelated inputs $A_{1} $ and $A_{2} $. The local input is denoted by $A_{i} $, and the local outputs and environments are denoted by $B_{i} $ and $E_{i} $, respectively. The remote output with respect to local map ${\rm {\mathcal M}}_{i} $, $j\ne i$, is referred to as $B_{j} $. The inputs can convey classical or quantum information, both the same type. A local map ${\rm {\mathcal M}}_{i} $ can be decomposed into the local logical channel ${\rm {\mathcal N}}_{A_{i} B_{i} } $, which exists between the input $A_{i} $ and the output $B_{i} $, and the local complementary channel ${\rm {\mathcal N}}_{A_{i} E_{i} } $, which connects the input $A_{i} $ with the local environment state $E_{i} $. Both ${\rm {\mathcal N}}_{A_{i} B_{i} } $ and ${\rm {\mathcal N}}_{A_{i} E_{i} } $ are qubit maps. In particular, for modeling purposes, we also introduce a $C$ qubit state, which identifies the realizations of the two local maps ${\rm {\mathcal M}}_{1} $ and ${\rm {\mathcal M}}_{2} $ by qubit states $C\in \left\{{\left| 0 \right\rangle} ,{\left| 1 \right\rangle} \right\}$.

Let $p={\textstyle\frac{1}{2}} $ be the probability of each map. Assuming a fixed causality, system $C$ can be modeled as a $d=2$ dimensional system with density
\begin{equation} \label{1)} 
\rho _{C} ={\textstyle\frac{1}{2}} \left({\left| 0 \right\rangle} {\left\langle 0 \right|} +{\left| 1 \right\rangle} {\left\langle 1 \right|} \right).                                
\end{equation} 

If the causality is non-fixed between the two local maps ${\rm {\mathcal M}}_{1} $ and ${\rm {\mathcal M}}_{2} $, then $C$ can be characterized by the superposition qubit state $C={\left| + \right\rangle} ={\textstyle\frac{1}{\sqrt{2} }} \left({\left| 0 \right\rangle} +{\left| 1 \right\rangle} \right)$, leading to the density
\begin{equation} \label{ZEqnNum707760} 
\rho _{C} ={\textstyle\frac{1}{2}} \left({\left| 0 \right\rangle} {\left\langle 0 \right|} +{\left| 0 \right\rangle} {\left\langle 1 \right|} +{\left| 1 \right\rangle} {\left\langle 0 \right|} +{\left| 1 \right\rangle} {\left\langle 1 \right|} \right).                         
\end{equation} 

Our investigation here is that the quantum gravity environment ${\rm {\mathcal G}}_{E} $, which models the space-time geometry (Theorem 3 will reveal that the local environment states also must be qubit states), does exactly the same controlling mechanism as a superposition qubit state $C={\left| + \right\rangle} $. However, there is a fundamental difference between systems $C$ and ${\rm {\mathcal G}}_{E} $. Although $C$ can be modeled by as a separable qubit state, in the quantum gravity setting, ${\rm {\mathcal G}}_{E} $ is a subsystem of an entangled tripartite system $\rho _{{\rm {\mathcal G}}_{E} E_{i} B_{j} } $, where the quantum gravity environment ${\rm {\mathcal G}}_{E} $ is entangled with the two-qubit system $E_{i} B_{j} $ via partition ${\rm {\mathcal G}}_{E} -E_{i} B_{j} $, that is, the system ${\rm {\mathcal G}}_{E} $ is non-separable from $E_{i} B_{j} $. This injects a fundamental difference between our model and that studied in \cite{ref11} because, in our model, the simultaneous realizations of the local maps ${\rm {\mathcal M}}_{1} $ and ${\rm {\mathcal M}}_{2} $ are a consequence of the entangled tripartite qubit system $\rho _{{\rm {\mathcal G}}_{E} E_{i} B_{j} } $ and a dedicated qubit superposition control system $C$ does not exist. 

However, the control state formalism $C={\left| + \right\rangle} $ still can be utilized to model the vanishing causality of the ${\rm {\mathcal M}}_{1} $ and ${\rm {\mathcal M}}_{2} $ local maps in our model, as it will be shown in \sref{sec4}.

Specifically, taking the Kraus operators of the local channels ${\rm {\mathcal N}}_{A_{1} E_{1} } $and ${\rm {\mathcal N}}_{A_{2} B_{2} } $ of maps ${\rm {\mathcal M}}_{1} $ and ${\rm {\mathcal M}}_{2} $
\begin{equation} \label{3)} 
{\rm {\mathcal N}}_{A_{1} E_{1} } \left(\rho \right)=\sum _{i}A_{i}^{A_{1} E_{1} }  \rho \left(A_{i}^{A_{1} E_{1} } \right)^{\dag } ,                           
\end{equation} 
\begin{equation} \label{4)} 
{\rm {\mathcal N}}_{A_{2} B_{2} } \left(\rho \right)=\sum _{j}A_{j}^{A_{2} B_{2} }  \rho \left(A_{j}^{A_{2} B_{2} } \right)^{\dag } ,                           
\end{equation} 
a CPTP map, ${\rm {\mathcal M}}_{{\rm {\mathcal G}}} $, can be introduced that describes the parallel realizations of the local channels ${\rm {\mathcal N}}_{A_{1} E_{1} } $ and ${\rm {\mathcal N}}_{A_{2} B_{2} } $. This map is defined as follows:
\begin{equation} \label{ZEqnNum710204} 
{\rm {\mathcal M}}_{{\rm {\mathcal G}}} \left(\rho \right)=\sum _{i}A_{i}^{{\rm {\mathcal G}}} \rho \left(A_{i}^{{\rm {\mathcal G}}} \right)^{\dag }  ,                              
\end{equation} 
where the Kraus operator $A_{i}^{{\rm {\mathcal G}}} $ is expressed as
\begin{equation} \label{6)} 
A_{i}^{{\rm {\mathcal G}}} ={\left| 0 \right\rangle} {\left\langle 0 \right|} \otimes A_{i}^{A_{1} E_{1} } \otimes A_{j}^{A_{2} B_{2} } +{\left| 1 \right\rangle} {\left\langle 1 \right|} \otimes A_{j}^{A_{2} B_{2} } \otimes A_{i}^{A_{1} E_{1} } .                
\end{equation} 

The local environment state and remote outputs $E_{1} $ and $B_{2} $ of ${\rm {\mathcal M}}_{1} $ and ${\rm {\mathcal M}}_{2} $ are entangled with the quantum gravity environment state ${\rm {\mathcal G}}_{E} $, formulating a mixed tripartite entangled qubit system $\rho _{{\rm {\mathcal G}}_{E} E_{1} B_{2} } $, in which $E_{1} $ is separable from ${\rm {\mathcal G}}_{E} B_{2} $, $B_{2} $ is separable from ${\rm {\mathcal G}}_{E} E_{1} $, and ${\rm {\mathcal G}}_{E} $ is entangled with $E_{1} B_{2} $. Together with the local environment $E_{2} $ and remote output $B_{1} $, systems $\rho _{{\rm {\mathcal G}}_{E} E_{1} B_{2} } $ and $\rho _{{\rm {\mathcal G}}_{E} E_{1} B_{2} } $ formulate the density matrix
\begin{equation} \label{7)} 
\rho ={\textstyle\frac{1}{2}} \rho _{{\rm {\mathcal G}}_{E} E_{1} B_{2} } +{\textstyle\frac{1}{2}} \rho _{{\rm {\mathcal G}}_{E} E_{2} B_{1} } .                              
\end{equation} 

Focusing on the tripartite system $\rho _{{\rm {\mathcal G}}_{E} E_{1} B_{2} } $ throughout, the following conditions hold for the partitions ${\rm {\mathcal G}}_{E} -E_{1} B_{2} $, $E_{1} -{\rm {\mathcal G}}_{E} B_{2} $, and $B_{2} -{\rm {\mathcal G}}_{E} E_{1} $ (Note: see also \cite{ref36}, and the references within for further details). 

Because the local subsystems $E_{1} $ and $B_{2} $ are separable from the partitions ${\rm {\mathcal G}}_{E} B_{2} $ and ${\rm {\mathcal G}}_{E} E_{1} $, in this tripartite system, only the quantum gravity environment ${\rm {\mathcal G}}_{E} $ can be entangled with $E_{1} B_{2} $, and all other partitions are separable with respect to $E_{1} $ and $B_{2} $. From these, it clearly follows that the partitions $E_{1} -{\rm {\mathcal G}}_{E} B_{2} $ and $B_{2} -{\rm {\mathcal G}}_{E} E_{1} $ are separable, and ${\rm {\mathcal G}}_{E} -E_{1} B_{2} $ is entangled \cite{ref36}.

Without loss of generality, we define a tripartite qubit system that simultaneously satisfies these conditions as
\begin{equation} \label{ZEqnNum477720} 
\rho _{{\rm {\mathcal G}}_{E} E_{1} B_{2} } =\Omega \cdot \xi +\left(1-\Omega \right)\chi ,                             
\end{equation} 
where 
\begin{equation} \label{ZEqnNum107175} 
\Omega \le {\textstyle\frac{1}{3}} .                                    
\end{equation} 
We further evaluate $\rho _{{\rm {\mathcal G}}_{E} E_{1} B_{2} } $ in \eqref{ZEqnNum477720} as
\begin{equation} \label{ZEqnNum436494}
\begin{split}
   {{\rho }_{{{\mathcal{G}}_{E}}{{E}_{1}}{{B}_{2}}}}=&\text{ }\tfrac{1}{2}\Omega \left( \left| 000 \right\rangle \left\langle  000 \right|+\left| 110 \right\rangle \left\langle  110 \right| \right) \\ 
 & +\text{ }\tfrac{1}{2}\left( \tfrac{1}{2}-\tfrac{1}{2}\Omega  \right)\left( \begin{split}
  & \left| 000 \right\rangle \left\langle  110 \right|+\left| 110 \right\rangle \left\langle  000 \right|+\left| 001 \right\rangle \left\langle  001 \right| \\ 
 & +\left| 011 \right\rangle \left\langle  011 \right|+\left| 101 \right\rangle \left\langle  101 \right|+\left| 111 \right\rangle \left\langle  111 \right| \\ 
\end{split} \right),  
\end{split}
\end{equation} 
where subsystem $\rho _{{\rm {\mathcal G}}_{E} E_{1} } $ is a separable Bell diagonal state, which can be expressed as
\begin{equation}  \label{11)}
\begin{split}
   {{\rho }_{{{\mathcal{G}}_{E}}{{E}_{1}}}}=&\tfrac{1}{2}\left( \tfrac{1}{2}+\tfrac{1}{2}\Omega  \right)\left( \left| 00 \right\rangle \left\langle  00 \right|+\left| 11 \right\rangle \left\langle  11 \right| \right) \\ 
 & +\tfrac{1}{2}\left( \tfrac{1}{2}-\tfrac{1}{2}\Omega  \right)\left( \left| 00 \right\rangle \left\langle  11 \right|+\left| 11 \right\rangle \left\langle  00 \right| \right) \\ 
 & +\tfrac{1}{2}\left( \tfrac{1}{2}-\tfrac{1}{2}\Omega  \right)\left( \left| 01 \right\rangle \left\langle  01 \right|+\left| 10 \right\rangle \left\langle  10 \right| \right).  
\end{split}
\end{equation} 
The density matrix in \eqref{ZEqnNum436494} can be rewritten as
\begin{equation}  \label{12)}
{{\rho }_{{{\mathcal{G}}_{E}}{{E}_{1}}{{B}_{2}}}}=\tfrac{1}{2}\left( \begin{matrix}
   \Omega  & 0 & 0 & 0 & 0 & 0 & \tfrac{1}{2}-\tfrac{1}{2}\Omega  & 0  \\
   0 & \tfrac{1}{2}-\tfrac{1}{2}\Omega  & 0 & 0 & 0 & 0 & 0 & 0  \\
   0 & 0 & 0 & 0 & 0 & 0 & 0 & 0  \\
   0 & 0 & 0 & \tfrac{1}{2}-\tfrac{1}{2}\Omega  & 0 & 0 & 0 & 0  \\
   0 & 0 & 0 & 0 & 0 & 0 & 0 & 0  \\
   0 & 0 & 0 & 0 & 0 & \tfrac{1}{2}-\tfrac{1}{2}\Omega  & 0 & 0  \\
   \tfrac{1}{2}-\tfrac{1}{2}\Omega  & 0 & 0 & 0 & 0 & 0 & \Omega  & 0  \\
   0 & 0 & 0 & 0 & 0 & 0 & 0 & \tfrac{1}{2}-\tfrac{1}{2}\Omega   \\
\end{matrix} \right),
\end{equation} 
whereas $\rho _{{\rm {\mathcal G}}_{E} E_{1} } $ can be expressed in as
\begin{equation} \label{ZEqnNum503683}
{{\rho }_{{{\mathcal{G}}_{E}}{{E}_{1}}}}=\tfrac{1}{2}\left( \begin{matrix}
   \tfrac{1}{2}+\tfrac{1}{2}\Omega  & 0 & 0 & \tfrac{1}{2}-\tfrac{1}{2}\Omega   \\
   0 & \tfrac{1}{2}-\tfrac{1}{2}\Omega  & 0 & 0  \\
   0 & 0 & \tfrac{1}{2}-\tfrac{1}{2}\Omega  & 0  \\
   \tfrac{1}{2}-\tfrac{1}{2}\Omega  & 0 & 0 & \tfrac{1}{2}+\tfrac{1}{2}\Omega   \\
\end{matrix} \right).
\end{equation} 
These will be referred via the partitions ${\rm {\mathcal G}}_{E} E_{1} $, ${\rm {\mathcal G}}_{E} B_{2} $, and ${\rm {\mathcal G}}_{E} -E_{1} B_{2} $ of $\rho _{{\rm {\mathcal G}}_{E} E_{1} B_{2} } $, respectively. In particular, for $\Omega \le {\textstyle\frac{1}{3}} $, the subsystems $\rho _{{\rm {\mathcal G}}_{E} E_{1} } $, $\rho _{{\rm {\mathcal G}}_{E} B_{2} } $, and $\rho _{E_{1} B_{2} } $ remain separable, while $\rho _{{\rm {\mathcal G}}_{E} } $ is entangled with $\rho _{E_{1} B_{2} } $; thus, it straightforwardly follows that the system of \eqref{ZEqnNum436494} can be used in the remaining part of the proof.

The separability conditions can be checked by taking the partial transposes $\left(\rho _{{\rm {\mathcal G}}_{E} E_{1} } \right)^{T_{{\rm {\mathcal G}}_{E} } } $, $\left(\rho _{{\rm {\mathcal G}}_{E} E_{1} } \right)^{T_{E_{1} } } $, $\left(\rho _{{\rm {\mathcal G}}_{E} E_{1} B_{2} } \right)^{T_{E_{1} } } $, and $\left(\rho _{{\rm {\mathcal G}}_{E} E_{1} B_{2} } \right)^{T_{B_{2} } } $ of $\rho _{{\rm {\mathcal G}}_{E} E_{1} B_{2} } $. 

The positivity of $\left(\rho _{{\rm {\mathcal G}}_{E} E_{1} } \right)^{T_{{\rm {\mathcal G}}_{E} } } $ and $\left(\rho _{{\rm {\mathcal G}}_{E} E_{1} } \right)^{T_{E_{1} } } $ trivially follows from \eqref{ZEqnNum503683} because $\rho _{{\rm {\mathcal G}}_{E} E_{1} } $ is a separable Bell diagonal state.

In particular, we will show the partial transpose of $\rho _{{\rm {\mathcal G}}_{E} E_{1} B_{2} } $ with respect to $B_{2} $, which can be expressed as follows:
\begin{equation} \label{ZEqnNum573970}
{{\left( {{\rho }_{{{\mathcal{G}}_{E}}{{E}_{1}}{{B}_{2}}}} \right)}^{{{T}_{{{B}_{2}}}}}}=\tfrac{1}{2}\left( \begin{matrix}
   \Omega  & 0 & 0 & 0 & 0 & 0 & 0 & 0  \\
   0 & \tfrac{1}{2}-\tfrac{1}{2}\Omega  & 0 & 0 & 0 & 0 & \tfrac{1}{2}-\tfrac{1}{2}\Omega  & 0  \\
   0 & 0 & 0 & 0 & 0 & 0 & 0 & 0  \\
   0 & 0 & 0 & \tfrac{1}{2}-\tfrac{1}{2}\Omega  & 0 & 0 & 0 & 0  \\
   0 & 0 & 0 & 0 & 0 & 0 & 0 & 0  \\
   0 & 0 & 0 & 0 & 0 & \tfrac{1}{2}-\tfrac{1}{2}\Omega  & 0 & 0  \\
   0 & \tfrac{1}{2}-\tfrac{1}{2}\Omega  & 0 & 0 & 0 & 0 & \Omega  & 0  \\
   0 & 0 & 0 & 0 & 0 & 0 & 0 & \tfrac{1}{2}-\tfrac{1}{2}\Omega   \\
\end{matrix} \right).
\end{equation}
This partial transpose is non-negative; hence,
\begin{equation} \label{15)} 
\left(\rho _{{\rm {\mathcal G}}_{E} E_{1} B_{2} } \right)^{T_{B_{2} } } \ge 0,                               
\end{equation} 
and similarly, with respect to $E_{1} $,
\begin{equation} \label{ZEqnNum455673} 
\left(\rho _{{\rm {\mathcal G}}_{E} E_{1} B_{2} } \right)^{T_{E_{1} } } \ge 0 .
\end{equation} 

Tracing out $B_{2} $ from $\rho _{{\rm {\mathcal G}}_{E} E_{1} B_{2} } $, one can check easily that the partial transpose of the resulting matrix $Tr_{B_{2} } \left(\rho _{{\rm {\mathcal G}}_{E} E_{1} B_{2} } \right)$ with respect to ${\rm {\mathcal G}}_{E} $ and $E_{2} $ is positive because $\left(\rho _{{\rm {\mathcal G}}_{E} E_{1} } \right)^{T_{{\rm {\mathcal G}}_{E} } } \ge 0$ and $\left(\rho _{{\rm {\mathcal G}}_{E} E_{1} } \right)^{T_{E_{1} } } \ge 0$.

Specifically, the partial transpose of $\rho _{{\rm {\mathcal G}}_{E} E_{1} B_{2} } $ with respect to ${\rm {\mathcal G}}_{E} $ is negative; hence,
\begin{equation} \label{17)} 
\left(\rho _{{\rm {\mathcal G}}_{E} E_{1} B_{2} } \right)^{T_{{\rm {\mathcal G}}_{E} } } <0,                              
\end{equation} 
which immediately proves that the quantum gravity environment ${\rm {\mathcal G}}_{E} $ (the space-time geometry) is entangled with $E_{1} B_{2} $.

The entangled structure of quantum gravity environment ${\rm {\mathcal G}}_{E} $ is depicted in \fref{fig1}. The information transmission is realized through the partition ${\rm {\mathcal G}}_{E} -E_{i} B_{j} $.

 \begin{center}
\begin{figure*}[h!]
\begin{center}
\includegraphics[angle = 0,width=1\linewidth]{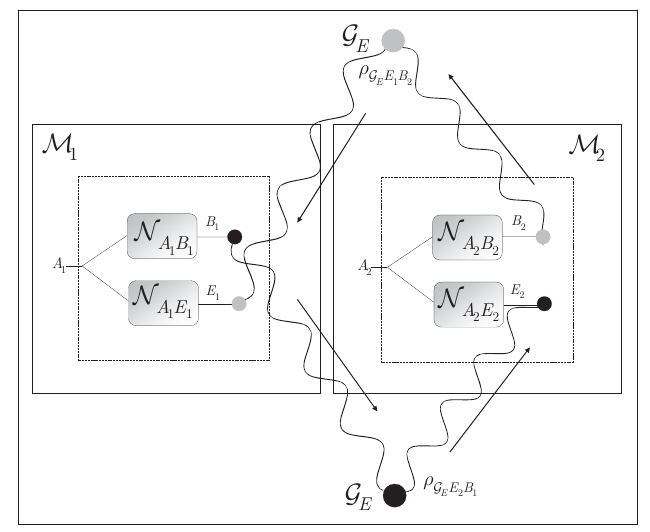}
\caption{The density matrix $\rho ={\textstyle\frac{1}{2}} \rho _{{\rm {\mathcal G}}_{E} E_{1} B_{2} } +{\textstyle\frac{1}{2}} \rho _{{\rm {\mathcal G}}_{E} E_{2} B_{1} } $. The local environment state $E_{i} $ and the remote output $B_{j} $ of ${\rm {\mathcal M}}_{1} $ and ${\rm {\mathcal M}}_{2} $ are entangled with the quantum gravity environment state ${\rm {\mathcal G}}_{E} $, via the partition ${\rm {\mathcal G}}_{E} -E_{i} B_{j} $. The entanglement between the local environments and the quantum gravity environment (or space-time geometry) allows the parties to simulate locally the remote outputs from their local environment. (The wavy lines illustrate the entanglement; the arrow refers to the direction of the information flow.)} 
 \label{fig1}
 \end{center}
\end{figure*}
\end{center}

From the map ${\rm {\mathcal M}}_{{\rm {\mathcal G}}} $ of \eqref{ZEqnNum710204}, it follows that the entangled structure of the density $\rho $ leads to a non-fixed causality between the local maps ${\rm {\mathcal M}}_{1} $ and ${\rm {\mathcal M}}_{2} $. The proof is concluded here. 
\end{proof}

Note that the strength of the correlation of the local environment state $E_{i} $ and the remote output $B_{j} $, $i\ne j$ can be characterized by the amount of information that is transferred through partitions ${\rm {\mathcal G}}_{E} E_{1} $ and ${\rm {\mathcal G}}_{E} E_{2} $. These questions, along with the information transmission capabilities of the quantum gravity environment, will be revealed next.

\begin{theorem}
(The information resource-pool property of quantum gravity). Quantum gravity acts as a background noise in form of a noisy CPTP map ${\rm {\mathcal D}}^{E_{i} \to B_{j} } $ on the local environment state $E_{i} $, which allows the parties to simulate the remote output $B_{j} $ as $B_{j} =E_{i} \circ {\rm {\mathcal D}}^{E_{i} \to B_{j} } $ with probability $p>{\textstyle\frac{1}{2}} $. The quantum gravity environment is an information resource-pool for the local parties.
\end{theorem}
\begin{proof}
Theorem 1 has revealed that, in the quantum gravity space, the local environment $E_{i} $ and the remote output $B_{j} $ of the local maps ${\rm {\mathcal M}}_{1} $ and ${\rm {\mathcal M}}_{2} $ together with the quantum gravity environment ${\rm {\mathcal G}}_{E} $ formulate an entangled tripartite qubit structure. We step forward from this point and show that the entangled ${\rm {\mathcal G}}_{E} -E_{i} B_{j} $ structure allows the local parties to simulate the remote output $B_{j} $ from the local environment $E_{i} $ with probability $p>{\textstyle\frac{1}{2}} $, above the classical limit $p={\textstyle\frac{1}{2}} $ (i.e., a coin tossing), which is precisely the case in a fixed causality structure where the local parties are independent \cite{ref10}.

The quantum gravity setting allows the parties with a probability $p$ to simulate the remote output from the local environment state through the local degrading map, in which the degrading map is a consequence of the quantum gravity environment. The remote output locally will be simulated from the local environment $E_{i} $ via the local CPTP map ${\rm {\mathcal D}}^{E_{i} \to B_{j} } $, the process of which is called remote simulation. It means that Alice can simulate $B_{2} $ from her local environment state $E_{1} $ as $B_{2} =E_{1} \circ {\rm {\mathcal D}}^{E_{1} \to B_{2} } $, and vice versa, Bob can simulate Alice's output $B_{1} $ as $B_{1} =E_{2} \circ {\rm {\mathcal D}}^{E_{2} \to B_{1} } $ (Note: The notation $\circ $ stands for the simulation, and the local degrading map ${\rm {\mathcal D}}^{E_{i} \to B_{j} } $ will be used in the right-hand side in the equations throughout.)

However, in the quantum gravity scenario, the information transmission through the partitions cannot be described by an ideal (i.e., noiseless) map; thus, the local degrading map ${\rm {\mathcal D}}^{E_{i} \to B_{j} } $ can be applied only with success probability $p$. Thus, the remote simulation is a probabilistic process. If the $I$ identity map is realized on $E_{i} $, then the remote simulation is not possible from $E_{i} $. This outcome has probability $1-p$.

In particular, the probabilistic remote simulation process can be characterized by a CPTP map ${\rm {\mathcal M}}_{{\rm {\mathcal D}}} $, defined as
\begin{equation} \label{ZEqnNum335498} 
{\rm {\mathcal M}}_{{\rm {\mathcal D}}} =p{\rm {\mathcal D}}^{E_{i} \to B_{j} } +\left(1-p\right)I, 
\end{equation} 
and the output of this map is as follows:
\begin{equation} \label{19)}
\begin{split}
   {{{{B}'_{j}}}}&={{E}_{i}}\circ {{\mathcal{M}}_{\mathcal{D}}} \\ 
 & ={{E}_{i}}\circ \left( p{{\mathcal{D}}^{{{E}_{i}}\to {{B}_{j}}}}+\left( 1-p \right)I \right) \\ 
 & =p{{B}_{j}}+\left( 1-p \right){{E}_{i}}.  
\end{split}
\end{equation} 
It is trivial that if the parties have no information about each other, then the remote output $B_{j} $ can be simulated from the local environment $E_{i} $ only with probability $p={\textstyle\frac{1}{2}} $; hence,
\begin{equation} \label{ZEqnNum375726} 
{\rm {\mathcal M}}_{{\rm {\mathcal D}}} ={\textstyle\frac{1}{2}} \left({\rm {\mathcal D}}^{E_{i} \to B_{j} } +I\right) 
\end{equation} 
and
\begin{equation} \label{21)} 
B'_{j} ={\textstyle\frac{1}{2}} \left(B_{j} +E_{i} \right). 
\end{equation} 
This is precisely the case in a standard scenario, where the quantum gravity effects are not present. The situation changes if we step into the quantum gravity space, which leads to success probability $p>{\textstyle\frac{1}{2}} $. To see it, we demonstrate this statement by assuming a case when both local CPTP maps ${\rm {\mathcal M}}_{1} $ and ${\rm {\mathcal M}}_{2} $ are the so-called entanglement-breaking channels.

The Kraus representation of the ${\rm {\mathcal M}}_{1} $ entanglement-breaking channel is evaluated as
\begin{equation} \label{22)} 
{{\mathcal{M}}_{1}}\left( {{\rho }_{A{A}'}} \right)=I\left( {{\rho }_{A}} \right)\otimes {{\mathcal{M}}_{1}}\left( {{\rho }_{{{A}'}}} \right)=\sum\limits_{i}{N_{i}^{\left( {{A}'} \right)}}{{\rho }_{A}}_{{{A}'}}N_{i}^{\left( {{A}'} \right)\dagger },
\end{equation} 
where $\rho _{AA'} $ refers to an entangled input system, and
\begin{equation} \label{23)} 
N_{i}^{\left(A'\right)} =I_{A} \otimes {\left| \xi _{i}  \right\rangle} _{A''} {\left\langle \varsigma  \right|} _{A'} , 
\end{equation} 
where $A'$ and $A''$ refer to the input and output systems, and the Kraus-operators $N_{i}^{\left(A'\right)} $ are unit rank. The sets $\left\{{\left| \xi _{i}  \right\rangle} _{A''} \right\}$ and $\left\{{\left| \varsigma  \right\rangle} _{A'} \right\}$ each do not necessarily form an orthonormal set.

Thus, for an entangled input $A'_{i} $ of an entanglement-breaking channel ${\rm {\mathcal N}}_{A_{i} B_{i} } $, it will destroy every entanglement on its local output $B_{i} $. Assuming a maximally entangled input system ${\left| \Psi  \right\rangle} _{AA'} ={\textstyle\frac{1}{\sqrt{d} }} \sum _{i=0}^{d-1}{\left| i \right\rangle} _{A}  {\left| i \right\rangle} _{A'} $, the output of ${\rm {\mathcal M}}_{1} $ can be expressed as follows:
\begin{equation} \label{24)} 
{\rm {\mathcal M}}_{1} \left({\left| \Psi  \right\rangle} {\left\langle \Psi  \right|} _{AA'} \right)=\sum _{x}p_{X} \left(x\right) \rho _{x}^{A} \otimes \rho _{x}^{B} ,                       
\end{equation} 
where $p_{X} \left(x\right)$ represents an arbitrary probability distribution, and $\rho _{x}^{A} $ and $\rho _{x}^{B} $ are the separable density matrices of the output system. The logical channel ${\rm {\mathcal N}}_{A_{i} B_{i} } $ performs a complete von Neumann measurement on its input system $\rho $ and outputs $\sigma {\rm }={\rm {\mathcal N}}_{EB} \left(\rho \right)$; hence, ${\rm {\mathcal N}}_{A_{i} B_{i} } $ is expressed as
\begin{equation} \label{25)} 
{\rm {\mathcal N}}_{A_{i} B_{i} } \left(\rho \right)=\sum _{x}\text{Tr}\left\{\Pi _{x} \rho \right\} \sigma _{x} , 
\end{equation} 
where $\left\{\Pi _{x} \right\}$ represents a positive operator valued measure (POVM) on $\rho $, and $\sigma _{x} $ is the output density matrix of the channel \cite{ref42}. The local ${\rm {\mathcal N}}_{A_{i} B_{i} } $ further can be decomposed into the CPTP map ${\rm {\mathcal N}}_{A_{i} B_{i} }^{1} $, a measurement operator $\left\{\Pi _{x} \right\}$, and a second map ${\rm {\mathcal N}}_{A_{i} B_{i} }^{2} $, which outputs the density matrix $\sigma _{x} $, together called conditional state preparation:
\begin{equation} \label{26)} 
{\rm {\mathcal N}}_{A_{i} B_{i} } ={\rm {\mathcal N}}_{A_{i} B_{i} }^{1} \circ \Pi \circ {\rm {\mathcal N}}_{A_{i} B_{i} }^{2} ,                     
\end{equation} 
where ${\rm {\mathcal N}}_{A_{i} B_{i} }^{1} =I$ and ${\rm {\mathcal N}}_{A_{i} B_{i} }^{2} =I$.

Introducing the notation $\Pi ^{X} $ for the $X$-basis, where $X$ refers to the Pauli operator $X$, and $\Pi ^{Z} $ for the $Z$-basis, where $Z$ refers to the Pauli operator $Z$, let the local ${\rm {\mathcal N}}_{A_{i} B_{i} } $ channels be defined as follows:
\begin{equation} \label{ZEqnNum729880} 
{\rm {\mathcal N}}_{A_{1} B_{1} } =I\circ \Pi _{1} \circ I,                                
\end{equation} 
where
\begin{equation} \label{28)} 
\Pi _{1} ={\textstyle\frac{1}{2}} \Pi ^{X} +{\textstyle\frac{1}{2}} \Pi ^{Z} ,                                
\end{equation} 
and 
\begin{equation} \label{29)} 
{\rm {\mathcal N}}_{A_{2} B_{2} } =I\circ \Pi _{2} \circ I,                                
\end{equation} 
where
\begin{equation} \label{30)} 
\Pi _{2} =\Pi ^{Z} . 
\end{equation} 
Let the local ${\rm {\mathcal D}}^{E_{i} \to B_{j} } $ maps of ${\rm {\mathcal M}}_{1} $ and ${\rm {\mathcal M}}_{2} $ be defined as follows:
\begin{equation} \label{ZEqnNum642978} 
{\rm {\mathcal D}}^{E_{1} \to B_{2} } =\Pi ^{Z}, 
\end{equation} 
and 
\begin{equation} \label{ZEqnNum463144} 
{\rm {\mathcal D}}^{E_{2} \to B_{1} } =\Pi ^{Z} .                                   
\end{equation} 
Thus, each ${\rm {\mathcal D}}^{E_{i} \to B_{j} } $ performs a projective measurement in the $Z$-basis on the local environment state $E_{i} $.

Using \eqref{ZEqnNum642978} and \eqref{ZEqnNum463144} along with the local channels ${\rm {\mathcal N}}_{A_{1} E_{1} } $ and ${\rm {\mathcal N}}_{A_{2} E_{2} } $, the remote outputs $B_{2} ,B_{1} $ are evaluated as
\begin{equation} \label{33)} 
B_{2} ={\rm {\mathcal N}}_{A_{1} E_{1} } \circ \Pi ^{Z} ,
\end{equation} 
and
\begin{equation} \label{34)} 
B_{1} ={\rm {\mathcal N}}_{A_{2} E_{2} } \circ \Pi ^{Z} .                                
\end{equation} 
For this setting, the state of $\rho _{{\rm {\mathcal G}}_{E} E_{i} B_{j} } $ is evaluated as follows:
\begin{equation} \label{ZEqnNum925947} 
{{\rho }_{{{\mathcal{G}}_{E}}{{E}_{i}}{{B}_{j}}}}=\left\{ \begin{matrix}
   {{\rho }_{{{\mathcal{G}}_{E}}{{E}_{2}}{{B}_{1}}}}\text{, if }{{\Pi }_{1}}={{\Pi }^{X}}  \\
   {{\rho }_{{{\mathcal{G}}_{E}}{{E}_{1}}{{B}_{2}}}}\text{, if }{{\Pi }_{1}}={{\Pi }^{Z}}  \\
\end{matrix} \right..
\end{equation} 
Thus, if $\Pi _{1} =\Pi ^{X} $, then Bob simulates Alice's output from his local environment $E_{2} $ through the partition ${\rm {\mathcal G}}_{E} -E_{2} B_{1} $ as $B_{1} ={\rm {\mathcal N}}_{A_{2} E_{2} } \circ \Pi ^{Z} $, whereas for $\Pi _{1} =\Pi ^{Z} $, Alice simulates Bob's output from $E_{1} $ via ${\rm {\mathcal G}}_{E} -E_{1} B_{2} $ as $B_{2} ={\rm {\mathcal N}}_{A_{1} E_{1} } \circ \Pi ^{Z} $.

The action of \eqref{ZEqnNum729880}-\eqref{ZEqnNum463144} can be rephrased by the process matrix formalism of \cite{ref10}, as follows. The process matrix $W^{B_{1} E_{1} B_{2} E_{2} } $ that describes the causality relations of the local maps ${\rm {\mathcal M}}_{1} $ and ${\rm {\mathcal M}}_{2} $ of $\rho _{{\rm {\mathcal G}}_{E} E_{i} B_{j} } $ in the quantum gravity scenario can be expressed as
\begin{equation} \label{36)} 
W^{B_{1} E_{1} B_{2} E_{2} } ={\textstyle\frac{1}{4}} \left(I^{B_{1} E_{1} B_{2} E_{2} } +{\textstyle\frac{1}{\sqrt{2} }} \left(Z^{E_{1} } X^{A_{1} } Z^{A_{2} } +Z^{E_{2} } Z^{A_{1} } \right)\right).
\end{equation} 
By applying the proof of Appendix E from \cite{ref10}, immediately yields that this process matrix identifies a causally non-separable process; and, the $p={\textstyle\frac{2+\sqrt{2} }{4}} $ success probability for the realization of the local degrading map ${\rm {\mathcal D}}^{E_{i} \to B_{j} } $ also straightforwardly follows for $W^{B_{1} E_{1} B_{2} E_{2} } $.

From these arguments, the main conclusion regarding the information resource-pool property of the quantum gravity environment can be derived. In the quantum gravity setting, the local map ${\rm {\mathcal D}}^{E_{i} \to B_{j} } $ can be realized with probability $p={\textstyle\frac{2+\sqrt{2} }{4}} $; hence, the local map ${\rm {\mathcal M}}_{{\rm {\mathcal D}}} $ from \eqref{ZEqnNum375726} can be rewritten as
\begin{equation} \label{37)} 
{\rm {\mathcal M}}_{{\rm {\mathcal D}}} ={\textstyle\frac{2+\sqrt{2} }{4}} {\rm {\mathcal D}}^{E_{i} \to B_{j} } +\left(1-{\textstyle\frac{2+\sqrt{2} }{4}} \right)I 
\end{equation} 
and
\begin{equation} \label{38)}
\begin{split}
   {{{{B}'_{j}}}}&={{E}_{i}}\circ {{\mathcal{M}}_{\mathcal{D}}} \\ 
 & ={{E}_{i}}\circ \left( \tfrac{2+\sqrt{2}}{4}{{\mathcal{D}}^{{{E}_{i}}\to {{B}_{j}}}}+\left( 1-\tfrac{2+\sqrt{2}}{4} \right)I \right) \\ 
 & =\tfrac{2+\sqrt{2}}{4}{{B}_{j}}+\left( 1-\tfrac{2+\sqrt{2}}{4} \right){{E}_{i}}.  
\end{split}
\end{equation} 
Thus, from the local environment $E_{i} $, the remote output $B_{j} $ can be simulated via the local map ${\rm {\mathcal M}}_{{\rm {\mathcal D}}} $ as $B_{j} =E_{i} \circ {\rm {\mathcal D}}^{E_{i} \to B_{j} } $ with probability $p>{\textstyle\frac{1}{2}} $. In particular, the quantum gravity environment acts as a noisy map on the local environment state and behaves as an information resource-pool for the local parties.

The model of remote simulation in the quantum gravity environment is summarized in \fref{fig2}.

 \begin{center}
\begin{figure*}[h!]
\begin{center}
\includegraphics[angle = 0,width=1\linewidth]{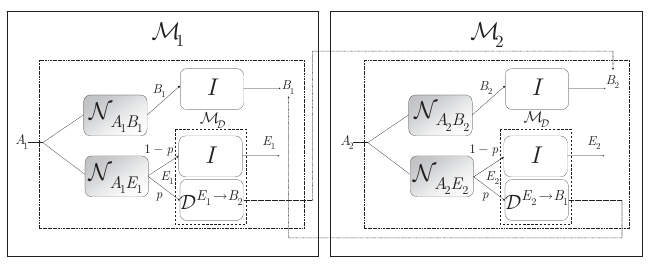}
\caption{The information resource-pool property of quantum gravity. The local CPTP maps ${\rm {\mathcal M}}_{1} $ and ${\rm {\mathcal M}}_{2} $ are independent, physically separated maps; the inputs $A_{1} $ and $A_{2} $ are uncorrelated variables conveying classical or quantum information; and ${\rm {\mathcal D}}^{E_{1} \to B_{2} } $ and ${\rm {\mathcal D}}^{E_{2} \to B_{1} } $ are local CPTP maps (called local degrading maps or background noise of quantum gravity). The local outputs and environment states are referred to as $B_{i} $,$E_{i} $, $i=1,2$, respectively. The quantum gravity setting allows the parties with a probability of $p>{\textstyle\frac{1}{2}} $ to simulate the remote output from the local environment state through the local degrading map ${\rm {\mathcal D}}^{E\to B} $. Alice can simulate $B_{2} $ from her local environment state $E_{1} $ as $B_{2} =E_{1} \circ {\rm {\mathcal D}}^{E_{1} \to B_{2} } $, whereas Bob can simulate Alice's output $B_{1} $ as $B_{1} =E_{2} \circ {\rm {\mathcal D}}^{E_{2} \to B_{1} } $. The quantum gravity acts as a noise on the local environments; thus, it behaves as an information resource-pool for the local parties about the remote CPTP maps.} 
 \label{fig2}
 \end{center}
\end{figure*}
\end{center}

These results confirm that, in the quantum gravity setting, there exists local independent CPTP maps, for which the local environments can be used to simulate the remote outputs with success probability $p>{\textstyle\frac{1}{2}} $. The quantum gravity environment, indeed, acts as an information resource-pool for the local parties. The proof is therefore concluded here.
\end{proof}

In Theorem 3, we reveal the structure of the quantum gravity channel that allows to model the quantum gravity space as an information transmission device between the $E_{i} $ local environment and the remote output $B_{j} $.
\begin{theorem}
(The structure of the quantum gravity channel) The local CPTP maps ${\rm {\mathcal N}}_{A_{i} B_{i} } $, ${\rm {\mathcal D}}^{E_{i} \to B_{j} } $, $i=1,2$, $j\ne i$, formulate the quantum gravity channel ${\rm {\mathcal M}}_{A_{i} B_{j} } $ with remote logical channel ${\rm {\mathcal N}}_{A_{i} B_{j} } ={\rm {\mathcal N}}_{A_{i} E_{i} } \circ {\rm {\mathcal D}}^{E_{i} \to B_{j} } $ and local complementary channel ${\rm {\mathcal N}}_{A_{i} E_{i} } $. The map ${\rm {\mathcal M}}_{A_{i} B_{j} } $ is anti-degradable, with local input $A_{i} $, remote output $B_{j} $, and local environment state $E_{i} $. 
\end{theorem}
\begin{proof}
In Theorem 2, we have seen that by exploiting the extra resources of quantum gravity, Alice can simulate Bob's output with probability $p>{\textstyle\frac{1}{2}} $, above the standard limit $p={\textstyle\frac{1}{2}} $. Here, we show that it leads to a well-defined channel structure---called the quantum gravity channel---between Alice and Bob. The causality structure of quantum gravity space-time geometry leads to an interesting configuration, namely, it brings alive a so-called remote simulation map, which acts locally at the parties, on their local environment states. 

The quantum gravity channel is referred by the CPTP map ${\rm {\mathcal M}}_{A_{i} B_{j} } $. The dimension of the local input $A_{i} $ of ${\rm {\mathcal M}}_{A_{i} B_{j} } $ is denoted by $d_{A_{i} } $, and the dimensions of the local environment $E_{i} $ and the remote output $B_{j} $ are referred as $d_{E_{i} } $ and $d_{B_{j} } $. The map ${\rm {\mathcal M}}_{A_{i} B_{j} } $ is decomposed into a logical channel ${\rm {\mathcal N}}_{A_{i} B_{j} } $ that exists between the local input $A_{i} $ and the remote output $B_{j} $, and into a local complementary channel ${\rm {\mathcal N}}_{A_{i} E_{i} } $, which exists between the local input $A_{i} $ and the local environment state $E_{i} $. The logical channel ${\rm {\mathcal N}}_{A_{i} B_{j} } $ is referred as the remote logical channel of ${\rm {\mathcal M}}_{A_{i} B_{j} } $ throughout, and it has the decomposition of ${\rm {\mathcal N}}_{A_{i} B_{j} } ={\rm {\mathcal N}}_{A_{i} E_{i} } \circ {\rm {\mathcal D}}^{E_{i} \to B_{j} } $; thus, this channel could exist only with probability $p$. 

Let us assume that the gravity channel ${\rm {\mathcal M}}_{A_{i} B_{j} } $ is an anti-degradable qubit channel. Then without loss of generality, the linear map of ${\rm {\mathcal M}}_{A_{i} B_{j} } :M_{2} \to M_{2} $ can be rewritten as
\begin{equation} \label{39)} 
{\rm {\mathcal M}}_{A_{i} B_{j} } :{\textstyle\frac{1}{2}} \left(I+\sum _{l}w_{k} \rho _{k}  \right)\to {\textstyle\frac{1}{2}} \left(I+\sum _{k}\left(t_{k} +\lambda _{k} w_{k} \right)\rho _{k}  \right),                  
\end{equation} 
where $t_{k} $ and $\lambda _{k} $ formulate the matrix $T_{{\rm {\mathcal M}}_{A_{i} B_{j} } } $ as
\begin{equation} \label{ZEqnNum936276}
{{T}_{{{\mathcal{M}}_{{{A}_{i}}{{B}_{j}}}}}}=\left( \begin{matrix}
   1 & 0 & 0 & 0  \\
   0 & {{\lambda }_{1}} & 0 & 0  \\
   0 & 0 & {{\lambda }_{2}} & 0  \\
   {{t}_{3}} & 0 & 0 & {{\lambda }_{3}}  \\
\end{matrix} \right).
\end{equation} 
From \eqref{ZEqnNum936276}, ${\rm {\mathcal M}}_{A_{i} B_{j} } $ can be rewritten as 
\begin{equation} \label{41)} 
\text{Tr}\rho _{l} {\rm {\mathcal M}}_{A_{i} B_{j} } \left(\rho _{k} \right).                               
\end{equation} 
For the input dimension $d_{A} $ of the qubit gravity channel ${\rm {\mathcal M}}_{A_{i} B_{j} } $ with local environment dimension $d_{E_{i} } =2$, a required condition on $d_{A_{i} } $ immediately follows from Theorem 4 of \cite{ref42}, namely, $d_{A_{i} } \le 3$. If $d_{A_{i} } =2$, then the remote output $B_{j} $ can be simulated from the local environment $E_{i} $, $i\ne j$ because the complementary channel ${\rm {\mathcal N}}_{A_{i} E_{i} } $ of ${\rm {\mathcal M}}_{A_{i} B_{j} } $ is degradable, whereas if $d_{A_{i} } =3$, then ${\rm {\mathcal N}}_{A_{i} E_{i} } $ is both degradable and anti-degradable.

Furthermore, because ${\rm {\mathcal M}}_{A_{i} B_{j} } $ is qubit channel, for the dimension $d_{B_{j} } $ of the remote output, the relation $d_{B_{j} } =2$ trivially follows. The condition $d_{E_{i} } =2$ on the Choi rank is satisfied only if
\begin{equation} \label{42)} 
\left(\lambda _{1} \pm \lambda _{2} \right)^{2} =\left(1\pm \lambda _{3} \right)^{2} -t_{3}^{2} ,                          
\end{equation} 
and
\begin{equation} \label{43)} 
\lambda _{3} =\lambda _{1} \lambda _{2} , 
\end{equation} 
\begin{equation} \label{44)} 
t_{3}^{2} =\left(1-\lambda _{1}^{2} \right)\left(1-\lambda _{2}^{2} \right), 
\end{equation} 
where $\left|\lambda _{i} \right|\le 1$.  

Introducing $u=v=\cos ^{-1} \left(\lambda _{1} \right)$, the matrix in \eqref{ZEqnNum936276} can be rewritten as 
\begin{equation}  \label{45)}
{{T}_{{{\mathcal{M}}_{{{A}_{i}}{{B}_{j}}}}}}=\left( \begin{matrix}
   1 & 0 & 0 & 0  \\
   0 & \cos u & 0 & 0  \\
   0 & 0 & \cos v & 0  \\
   \sin u\sin v & 0 & 0 & \cos u\cos v  \\
\end{matrix} \right),
\end{equation} 
where the anti-degradability of the qubit gravity channel ${\rm {\mathcal M}}_{A_{i} B_{j} } $ implies that
\begin{equation} \label{ZEqnNum296824} 
\sin u>\cos v,                               
\end{equation} 
which also follows from Theorem 5 of \cite{ref42}. The Kraus representation of ${\rm {\mathcal M}}_{A_{i} B_{j} } $ is ${\rm {\mathcal M}}_{A_{i} B_{j} } \left(\rho \right)=A_{+} \rho A_{+}^{\dag } +A_{-} \rho A_{-}^{\dag } $, where
\begin{equation}  \label{47)}
{{A}_{+}}=\cos \tfrac{1}{2}v\cos \tfrac{1}{2}uI+\sin \tfrac{1}{2}v\sin \tfrac{1}{2}uZ=\left( \begin{matrix}
   \cos \tfrac{1}{2}\left( v-u \right) & 0  \\
   0 & \cos \tfrac{1}{2}\left( u+u \right)  \\
\end{matrix} \right),
\end{equation} 
and
\begin{equation} \label{48)}
{{A}_{-}}=\sin \tfrac{1}{2}v\cos \tfrac{1}{2}uX-i\cos \tfrac{1}{2}v\sin \tfrac{1}{2}uY=\left( \begin{matrix}
   0 & \sin \tfrac{1}{2}\left( v-u \right)  \\
   \sin \tfrac{1}{2}\left( u+v \right) & 0  \\
\end{matrix} \right),
\end{equation} 
where $X$, $Y$, and $Z$ are the Pauli operators. One can get the condition $\left|\sin v\right|\ge \left|\cos u\right|$, which is analogous to \eqref{ZEqnNum296824}, however, in a slightly different form.

The structure of the quantum gravity channel ${\rm {\mathcal M}}_{A_{i} B_{j} } $ is summarized in \fref{fig3}.

 \begin{center}
\begin{figure*}[h!]
\begin{center}
\includegraphics[angle = 0,width=1\linewidth]{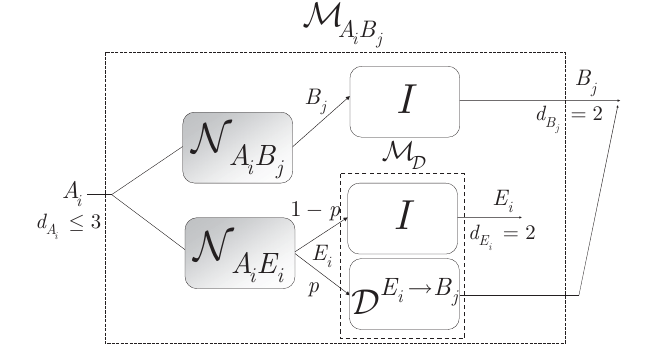}
\caption{The quantum gravity channel ${\rm {\mathcal M}}_{A_{i} B_{j} } $ with remote logical channel ${\rm {\mathcal N}}_{A_{i} B_{j} } $ and local complementary channel ${\rm {\mathcal N}}_{A_{i} E_{i} } $. The input of the channel is $A_{i} $, and the output is $B_{j} $, $i=1,2$, $i\ne j$. The ${\rm {\mathcal M}}_{A_{i} B_{j} } $ remote output channel exists with probability $p>{\textstyle\frac{1}{2}} $, and ${\rm {\mathcal M}}_{A_{i} B_{j} } $ is an anti-degradable map; thus, from the local environment state $E_{i} $, the remote output $B_{j} $ can be locally simulated by ${\rm {\mathcal D}}^{E_{i} \to B_{j} } $. In ${\rm {\mathcal M}}_{{\rm {\mathcal D}}} =p{\rm {\mathcal D}}^{E_{i} \to B_{j} } +\left(1-p\right)I$, the map ${\rm {\mathcal D}}^{E_{i} \to B_{j} } $ performs the so-called remote simulation.} 
 \label{fig3}
 \end{center}
\end{figure*}
\end{center}

Assume that there are two local maps, ${\rm {\mathcal M}}_{1} $ and ${\rm {\mathcal M}}_{2} $, in the system, with remote logical channels ${\rm {\mathcal N}}_{A_{1} B_{2} } $ and ${\rm {\mathcal N}}_{A_{2} B_{1} } $. Taking the superset ${\rm {\mathcal S}}$ of these gravity channels, the result is a convex set because ${\rm {\mathcal S}}$ formulates a supergravity channel as
\begin{equation} \label{49)} 
{\rm {\mathcal S}}={\textstyle\frac{1}{2}} {\rm {\mathcal N}}_{A_{1} B_{2} } \otimes {\left| 0 \right\rangle} {\left\langle 0 \right|} _{F} +{\textstyle\frac{1}{2}} {\rm {\mathcal N}}_{A_{2} B_{1} } \otimes {\left| 1 \right\rangle} {\left\langle 1 \right|} _{F} ,                      
\end{equation} 
with complementary channel 
\begin{equation} \label{50)} 
{\rm {\mathcal S}}^{C} ={\textstyle\frac{1}{2}} {\rm {\mathcal N}}_{A_{1} E_{1} } \otimes {\left| 0 \right\rangle} {\left\langle 0 \right|} _{G} +{\textstyle\frac{1}{2}} {\rm {\mathcal N}}_{A_{2} E_{2} } \otimes {\left| 1 \right\rangle} {\left\langle 1 \right|} _{G} ,                     
\end{equation} 
where $F$ and $G$ are elements of the Stinespring representation. 

From the set $\left\{A_{k}^{i} \right\}_{k} $ of Kraus operators of the remote simulation map ${\rm {\mathcal D}}^{E_{i} \to B_{j} } $, that is, ${\rm {\mathcal D}}^{E_{1} \to B_{2} } $ and ${\rm {\mathcal D}}^{E_{2} \to B_{1} } $, operator $A_{i} $ of ${\rm {\mathcal D}}^{E_{i} \to B_{j} } $ is as follows:
\begin{equation} \label{51)} 
A_{i} =A_{k}^{0} \otimes {\left| 0 \right\rangle} {\left\langle 0 \right|} +A_{k}^{1} \otimes {\left| 1 \right\rangle} {\left\langle 1 \right|} .                             
\end{equation} 
Applying ${\rm {\mathcal D}}^{E_{i} \to B_{j} } $ on ${\rm {\mathcal S}}^{C} $ results in 
\begin{equation} \label{ZEqnNum112352}
\begin{split}
   \mathcal{S}{{}^{C}}\circ {{\mathcal{D}}^{{{E}_{i}}\to {{B}_{j}}}}&=\tfrac{1}{2}\left( {{\mathcal{N}}_{{{A}_{1}}{{E}_{1}}}}\circ {{\mathcal{D}}^{{{E}_{1}}\to {{B}_{2}}}} \right)\otimes \left| 0 \right\rangle {{\left\langle  0 \right|}_{G}}+\tfrac{1}{2}\left( {{\mathcal{N}}_{{{A}_{2}}{{E}_{2}}}}\circ {{\mathcal{D}}^{{{E}_{2}}\to {{B}_{1}}}} \right)\otimes \left| 1 \right\rangle {{\left\langle  1 \right|}_{G}} \\ 
 & =\tfrac{1}{2}{{\mathcal{N}}_{{{A}_{1}}{{B}_{2}}}}\otimes \left| 0 \right\rangle {{\left\langle  0 \right|}_{F}}+\tfrac{1}{2}\left( {{\mathcal{N}}_{{{A}_{2}}{{B}_{1}}}} \right)\otimes \left| 1 \right\rangle {{\left\langle  1 \right|}_{F}} \\ 
 & =\mathcal{S}.  
\end{split}
\end{equation} 
Using Lemma 17 from \cite{ref42}, one can readily see that the super gravity channel ${\rm {\mathcal S}}$ is anti-degradable because applying map $\text{Tr}_{F} $ on \eqref{ZEqnNum112352} leads to
\begin{equation} \label{53)} 
\begin{split}
  & \text{Tr}_{F}\left( \mathcal{S}{{}^{C}}\circ {{\mathcal{D}}^{{{E}_{i}}\to {{B}_{j}}}} \right) \\ 
 & =\text{Tr}_{F}\left( \mathcal{S} \right) \\ 
 & =\text{Tr}_{F}\left( \tfrac{1}{2}{{\mathcal{N}}_{{{A}_{1}}{{B}_{2}}}}\otimes \left| 0 \right\rangle {{\left\langle  0 \right|}_{F}}+\tfrac{1}{2}\left( {{\mathcal{N}}_{{{A}_{2}}{{B}_{1}}}} \right)\otimes \left| 1 \right\rangle {{\left\langle  1 \right|}_{F}} \right) \\ 
 & =\tfrac{1}{2}{{\mathcal{N}}_{{{A}_{1}}{{B}_{2}}}}+\tfrac{1}{2}\left( {{\mathcal{N}}_{{{A}_{2}}{{B}_{1}}}} \right).  
\end{split}
\end{equation} 
These results conclude that the quantum gravity channel ${\rm {\mathcal M}}_{A_{i} B_{j} } $ is anti-degradable and allows the parties to perform the remote simulation of outputs $B_{j} $ from the local environment state $E_{i} $ by utilizing the map ${\rm {\mathcal D}}^{E_{i} \to B_{j} } $. This degrading map arises from the extra informational resource-pool property of quantum gravity, and the realization of this map is trivially not possible with probability $p>{\textstyle\frac{1}{2}} $ in the standard scenario, where the causality is fixed and non-vanishing. 
\end{proof}

\section{Information Transfer of Quantum Gravity}
\label{sec3}
\begin{theorem}
 (Information transfer of quantum gravity) The quantum gravity environment allows the transfer of classical and quantum information between the local maps ${\rm {\mathcal M}}_{1} $ and ${\rm {\mathcal M}}_{2} $. The information flow is realized through the quantum gravity environment via the partition ${\rm {\mathcal G}}_{E} -E_{i} B_{j} $ of the tripartite system $\rho _{{\rm {\mathcal G}}_{E} E_{i} B_{j} } $.
\end{theorem}
\begin{proof}
The correlation measure can be settled between subsystems ${\rm {\mathcal G}}_{E} E_{i} $ and ${\rm {\mathcal G}}_{E} B_{j} $. For simplicity, we will use ${\rm {\mathcal G}}_{E} E_{1} $ throughout to characterize exactly the information transmission between the local environment states and the quantum gravity environment state. We derive various correlation measures for the output system $\rho _{{\rm {\mathcal G}}_{E} E_{1} } $. 

Specifically, in this special quantum gravity communication scenario, Alice and Bob cannot transmit directly to each other any information. Instead of a direct signaling, the degraded local environment $B'_{2} =E_{1} \circ {\rm {\mathcal M}}_{{\rm {\mathcal D}}} $, see \eqref{ZEqnNum335498}, and the remote output $B_{2} $ will characterize the correlation between Alice and Bob's maps ${\rm {\mathcal M}}_{1} $ and ${\rm {\mathcal M}}_{2} $, despite the fact that all correlations are transmitted via the entangled quantum gravity environment. Thus, in fact, the communication is realized through the quantum gravity environment ${\rm {\mathcal G}}_{E} $, via ${\rm {\mathcal G}}_{E} E_{i} $ and ${\rm {\mathcal G}}_{E} B_{j} $. The entangled Hilbert space ${\rm {\mathcal G}}_{E} -E_{i} B_{j} $, in fact, acts as a communication channel. 

Assuming the case that Alice simulates Bob's output, we introduce the CPTP map
\begin{equation} \label{54)}
\begin{split}
   \mathcal{M}\left( {{B}_{2}} \right):&={{{{B}'_{2}}}} \\ 
 & ={{E}_{1}}\circ {{\mathcal{M}}_{\mathcal{D}}} \\ 
 & ={{E}_{1}}\circ \left( p{{\mathcal{D}}^{{{E}_{i}}\to {{B}_{j}}}}+\left( 1-p \right)I \right),  
\end{split}
\end{equation}
which gets the remote output $B_{2} $ as input and outputs Alice's noisy $B'_{2} =E_{1} \circ {\rm {\mathcal M}}_{{\rm {\mathcal D}}} $, see \eqref{ZEqnNum335498} and Theorem 3. Thus, it is a noisy evolution on Bob's ideal $B_{2} $ that results in $B'_{2} $. We step forward from the results of Theorem 3 to drive the information transmission capabilities of channel ${\rm {\mathcal M}}\left(B_{2} \right)$ by quantifying the amount of information that is conveyed by ${\rm {\mathcal G}}_{E} E_{1} $, using the system defined in \eqref{ZEqnNum477720}. Hence, the analysis will be made from Alice's viewpoint, via subsystem $\rho _{{\rm {\mathcal G}}_{E} E_{1} } $.

First, we rewrite the Bell-diagonal system $\rho _{{\rm {\mathcal G}}_{E} E_{1} } $ from \eqref{ZEqnNum503683} as
\begin{equation} \label{ZEqnNum900937} 
\rho _{{\rm {\mathcal G}}_{E} E_{1} } =\frac{1}{4} \left(I\otimes I+\textbf{r}\cdot \vec{\sigma }\otimes I+I\otimes \textbf{s}\cdot \vec{\sigma }+\sum _{i=1}^{3}c_{i} \sigma _{i} \otimes \sigma _{i}  \right),               
\end{equation} 
where $\textbf{r}$ and $\textbf{s}$ are the Bloch vectors, $\vec{\sigma }=\left[\sigma _{x} ,\sigma _{y} ,\sigma _{z} \right]$ with the Pauli matrices $\sigma _{i} $, and $c_{i} $ is the real parameter $-1\le c_{i} \le 1$ \cite{ref36,ref37,ref52}. For a Bell diagonal state $\textbf{r}=\textbf{s}=0$. For $\textbf{r}=\left(0,0,r\right)$ and $\textbf{s}=\left(0,0,s\right)$, the input state in \eqref{ZEqnNum900937} can be given in a matrix representation as follows: 
\begin{equation} \label{ZEqnNum906036}
{{\rho }_{{{\mathcal{G}}_{E}}{{E}_{1}}}}=\frac{1}{4}\left( \begin{matrix}
   1+r+s+{{c}_{3}} & 0 & 0 & {{c}_{1}}-{{c}_{2}}  \\
   0 & 1+r-s-{{c}_{3}} & {{c}_{1}}+{{c}_{2}} & 0  \\
   0 & {{c}_{1}}+{{c}_{2}} & 1-r+s-{{c}_{3}} & 0  \\
   {{c}_{1}}-{{c}_{2}} & 0 & 0 & 1-r-s+{{c}_{3}}  \\
\end{matrix} \right).
\end{equation}
The eigenvalues $u_{+} ,u_{-} ,v_{+} ,v_{-} $ of $\rho _{{\rm {\mathcal G}}_{E} E_{1} } $ are defined as
\begin{equation} \label{ZEqnNum632586} 
v_{+} =\frac{1}{4} \left(1-c_{3} +\sqrt{\left(r-s\right)^{2} +\left(c_{1} +c_{2} \right)^{2} } \right)\ge 0, 
\end{equation} 
\begin{equation} \label{58)} 
v_{-} =\frac{1}{4} \left(1-c_{3} -\sqrt{\left(r-s\right)^{2} +\left(c_{1} +c_{2} \right)^{2} } \right)\ge 0, 
\end{equation} 
\begin{equation} \label{ZEqnNum544980} 
u_{+} =\frac{1}{4} \left(1+c_{3} +\sqrt{\left(r+s\right)^{2} +\left(c_{1} -c_{2} \right)^{2} } \right)\ge 0, 
\end{equation} 
\begin{equation} \label{60)} 
u_{-} =\frac{1}{4} \left(1+c_{3} -\sqrt{\left(r+s\right)^{2} +\left(c_{1} -c_{2} \right)^{2} } \right)\ge 0. 
\end{equation} 
From these eigenvalues, the $-1\le c_{i} \le 1$ parameters of $\rho _{{\rm {\mathcal G}}_{E} E_{1} } $ can be expressed as
\begin{equation} \label{61)} 
c_{1} =\left(v_{+} -v_{-} \right), 
\end{equation} 
\begin{equation} \label{62)} 
c_{2} =-\left(v_{+} -v_{-} \right),                              
\end{equation} 
and
\begin{equation} \label{63)} 
c_{3} =1-2\cdot \left(v_{+} -v_{-} \right)=1+2\cdot c_{2} . 
\end{equation} 
For these parameters, the relations $\left|c_{1} \right|+\left|c_{2} \right|+\left|c_{3} \right|\le 1$ and $\max \left\{v_{+} ,v_{-} ,u_{+} ,u_{-} \right\}\le {\textstyle\frac{1}{2}} $ hold in \eqref{ZEqnNum906036}. Some trivial steps then straightforwardly yields that $\Omega $ can be expressed from the eigenvalues $v_{+} ,v_{-} $ as
\begin{equation} \label{64)} 
\Omega =1-2\left(v_{+} -v_{-} \right),                             
\end{equation} 
from which the correlations in $\rho _{{\rm {\mathcal G}}_{E} E_{1} } $ can be exactly determined in function of $\Omega $.

The $I\left(\rho _{{\rm {\mathcal G}}_{E} E_{1} } \right)$ mutual information function measures the total correlation in $\rho _{{\rm {\mathcal G}}_{E} E_{1} } $. The mutual information function of $\rho _{{\rm {\mathcal G}}_{E} E_{1} } $ can be expressed as follows:
\begin{equation} \label{ZEqnNum616334} 
I\left(\rho _{{\rm {\mathcal G}}_{E} E_{1} } \right)=S\left(\rho _{{\rm {\mathcal G}}_{E} } \right)+S\left(\rho _{E_{1} } \right)-S\left(\rho _{{\rm {\mathcal G}}_{E} E_{1} } \right).  
\end{equation} 
Using the eigenvalues of $\rho _{{\rm {\mathcal G}}_{E} E_{1} } $, $I\left(\rho _{{\rm {\mathcal G}}_{E} E_{1} } \right)$ can be rewritten as
\begin{equation} \label{66)} 
I\left(\rho _{{\rm {\mathcal G}}_{E} E_{1} } \right)=S\left(\rho _{{\rm {\mathcal G}}_{E} } \right)+S\left(\rho _{E_{1} } \right)+u_{+} \log _{2} u_{+} +u_{-} \log _{2} u_{-} +v_{+} \log _{2} v_{+} +v_{-} \log _{2} v_{-} ,  
\end{equation} 
where $S\left(\cdot \right)$ is the von Neumann entropy \cite{ref52} and
\begin{equation} \label{67)} 
S\left(\rho _{{\rm {\mathcal G}}_{E} } \right)=1-\frac{1}{2} \left(1-r\right)\log _{2} \left(1-r\right)-\frac{1}{2} \left(1+r\right)\log _{2} \left(1+r\right),                
\end{equation} 
\begin{equation} \label{68)} 
S\left(\rho _{E_{1} } \right)=1-\frac{1}{2} \left(1-s\right)\log _{2} \left(1-s\right)-\frac{1}{2} \left(1+s\right)\log _{2} \left(1+s\right).                 
\end{equation} 
The amount of purely classical correlation ${\rm {\mathcal C}}\left(\rho _{{\rm {\mathcal G}}_{E} E_{1} } \right)$ in $\rho _{{\rm {\mathcal G}}_{E} E_{1} } $ can be expressed as follows: 
\begin{equation} \label{ZEqnNum161225}
\begin{split}
   \mathcal{C}\left( {{\rho }_{{{\mathcal{G}}_{E}}{{E}_{1}}}} \right)&=S\left( {{\rho }_{{{E}_{1}}}} \right)-\tilde{S}\left( \left. {{E}_{1}} \right|{{\mathcal{G}}_{E}} \right) \\ 
 & =S\left( {{\rho }_{{{E}_{1}}}} \right)-\underset{{{E}_{k}}}{\mathop{\min }}\,\sum\limits_{k}{pkS\left( {{\sigma }_{\left. {{E}_{1}} \right|k}} \right)},  
\end{split}
\end{equation} 
where $\rho _{\left. E_{1} \right|k} ={\textstyle\frac{{\left\langle k \mathrel{\left| \vphantom{k \rho _{{\rm {\mathcal G}}_{E} } \left. \rho _{E_{1} } \right|k}\right.\kern-\nulldelimiterspace} \rho _{{\rm {\mathcal G}}_{E} } \left. \rho _{E_{1} } \right|k \right\rangle} }{{\left\langle k \mathrel{\left| \vphantom{k \rho _{{\rm {\mathcal G}}_{E} } k}\right.\kern-\nulldelimiterspace} \rho _{{\rm {\mathcal G}}_{E} } k \right\rangle} }} $ is the post-measurement state of $\rho _{E_{1} } $, the probability of result $k$ is $p_{k} =d{\left\langle k \mathrel{\left| \vphantom{k \rho _{{\rm {\mathcal G}}_{E} } k}\right.\kern-\nulldelimiterspace} \rho _{{\rm {\mathcal G}}_{E} } k \right\rangle} $, $d=2$ is the dimension of system $\rho _{{\rm {\mathcal G}}_{E} } $, and $q_{k} $ makes up a normalized probability distribution in the rank-one POVM elements $E_{k} =q_{k} {\left| k \right\rangle} {\left\langle k \right|} $. 

The purely classical correlation can also be expressed by the following formula: 
\begin{equation} \label{ZEqnNum825612} 
{\rm {\mathcal C}}\left(\rho _{{\rm {\mathcal G}}_{E} E_{1} } \right)=S\left(\rho _{{\rm {\mathcal G}}_{E} } \right)-\min \left\{f_{1} ,f_{2} ,f_{3} \right\}, 
\end{equation} 
where the functions $f_{1} ,f_{2} $, and $f_{3} $ are defined as follows \cite{ref36,ref37,ref52}:
\begin{equation} \label{71)}
\begin{split}
   {{f}_{1}}=&-\frac{1}{4}\left( 1+r+s+{{c}_{3}} \right){{\log }_{2}}\frac{1}{2\left( 1+s \right)}\left( 1+r+s+{{c}_{3}} \right) \\ 
 & -\frac{1}{4}\left( 1-r+s-{{c}_{3}} \right){{\log }_{2}}\frac{1}{2\left( 1+s \right)}\left( 1-r+s-{{c}_{3}} \right) \\ 
 & -\frac{1}{4}\left( 1+r-s-{{c}_{3}} \right){{\log }_{2}}\frac{1}{2\left( 1+s \right)}\left( 1+r-s-{{c}_{3}} \right) \\ 
 & -\frac{1}{4}\left( 1-r-s+{{c}_{3}} \right){{\log }_{2}}\frac{1}{2\left( 1+s \right)}\left( 1-r-s+{{c}_{3}} \right),  
\end{split}
\end{equation}
\begin{equation} \label{72)} 
f_{2} =1-\frac{1}{2} \left(1-\sqrt{r+c_{1}^{2} } \right)\log _{2} \left(1-\sqrt{r+c_{1}^{2} } \right)-\frac{1}{2} \left(1+\sqrt{r+c_{1}^{2} } \right)\log _{2} \left(1+\sqrt{r+c_{1}^{2} } \right),  
\end{equation} 
and 
\begin{equation} \label{73)} 
f_{3} =1-\frac{1}{2} \left(1-\sqrt{r+c_{2}^{2} } \right)\log _{2} \left(1-\sqrt{r+c_{2}^{2} } \right)-\frac{1}{2} \left(1+\sqrt{r+c_{2}^{2} } \right)\log _{2} \left(1+\sqrt{r+c_{2}^{2} } \right).  
\end{equation} 
As follows, the $C\left({\rm {\mathcal M}}\left(B_{2} \right)\right)$ classical capacity of channel ${\rm {\mathcal M}}\left(B_{2} \right)$ is
\begin{equation} \label{74)} 
C\left({\rm {\mathcal M}}\left(B_{2} \right)\right)=\mathop{\lim }\limits_{n\to \infty } \frac{1}{n} \mathop{\max }\limits_{\forall \rho _{{\rm {\mathcal G}}_{E} E_{1} } } I\left(\rho _{{\rm {\mathcal G}}_{E} E_{1} } \right).                      
\end{equation} 
From the mutual information $I\left(\rho _{{\rm {\mathcal G}}_{E} E_{1} } \right)$ and the classical correlation ${\rm {\mathcal C}}\left(\rho _{{\rm {\mathcal G}}_{E} E_{1} } \right)$, the ${\rm {\mathcal D}}\left(\rho _{{\rm {\mathcal G}}_{E} E_{1} } \right)$ quantum discord \cite{ref52} is as follows:
\begin{equation} \label{75)}
\begin{split}
   \mathcal{D}\left( {{\rho }_{{{\mathcal{G}}_{E}}{{E}_{1}}}} \right)=&I\left( {{\rho }_{{{\mathcal{G}}_{E}}{{E}_{1}}}} \right)-\mathcal{C}\left( {{\rho }_{{{\mathcal{G}}_{E}}{{E}_{1}}}} \right) \\ 
  =&S\left( {{\rho }_{{{\mathcal{G}}_{E}}}} \right)+S\left( {{\rho }_{{{E}_{1}}}} \right)+{{u}_{+}}{{\log }_{2}}{{u}_{+}}+{{u}_{-}}{{\log }_{2}}{{u}_{-}}+{{v}_{+}}{{\log }_{2}}{{v}_{+}}+{{v}_{-}}{{\log }_{2}}{{v}_{-}} \\ 
 & -\left( S\left( {{\rho }_{{{\mathcal{G}}_{E}}}} \right)-\min \left\{ {{f}_{1}},{{f}_{2}},{{f}_{3}} \right\} \right) \\ 
  =&S\left( {{\rho }_{{{E}_{1}}}} \right)+{{u}_{+}}{{\log }_{2}}{{u}_{+}}+{{u}_{-}}{{\log }_{2}}{{u}_{-}}+{{v}_{+}}{{\log }_{2}}{{v}_{+}}+{{v}_{-}}{{\log }_{2}}{{v}_{-}}+\min \left\{ {{f}_{1}},{{f}_{2}},{{f}_{3}} \right\}.  
\end{split}
\end{equation}
The $I_{coh} \left(\rho _{{\rm {\mathcal G}}_{E} E_{1} } \right)$ coherent information of $\rho _{{\rm {\mathcal G}}_{E} E_{1} } $ can be expressed as
\begin{equation}\label{76)}            
\begin{split}
   {{I}_{coh}}\left( {{\rho }_{{{\mathcal{G}}_{E}}{{E}_{1}}}} \right)&=\mathcal{D}\left( {{\rho }_{{{\mathcal{G}}_{E}}{{E}_{1}}}} \right)+\mathcal{C}\left( {{\rho }_{{{\mathcal{G}}_{E}}{{E}_{1}}}} \right)-1 \\ 
 & =I\left( {{\rho }_{{{\mathcal{G}}_{E}}{{E}_{1}}}} \right)-\mathcal{C}\left( {{\rho }_{{{\mathcal{G}}_{E}}{{E}_{1}}}} \right)+\mathcal{C}\left( {{\rho }_{{{\mathcal{G}}_{E}}{{E}_{1}}}} \right)-1 \\ 
 & =I\left( {{\rho }_{{{\mathcal{G}}_{E}}{{E}_{1}}}} \right)-1 \\ 
 & =S\left( {{\rho }_{{{\mathcal{G}}_{E}}}} \right)+S\left( {{\rho }_{{{E}_{1}}}} \right)+{{u}_{+}}{{\log }_{2}}{{u}_{+}}+{{u}_{-}}{{\log }_{2}}{{u}_{-}}+{{v}_{+}}{{\log }_{2}}{{v}_{+}}+{{v}_{-}}{{\log }_{2}}{{v}_{-}}-1.  
\end{split}
\end{equation}
The $Q\left({\rm {\mathcal M}}\left(B_{2} \right)\right)$ quantum capacity of ${\rm {\mathcal M}}\left(B_{2} \right)$ can be given as the maximization of the coherent information $I_{coh} \left(\rho _{{\rm {\mathcal G}}_{E} E_{1} } \right)$ of $\rho _{{\rm {\mathcal G}}_{E} E_{1} } $ as 
\begin{equation} \label{77)}
\begin{split}
   Q\left( \mathcal{M}\left( {{B}_{2}} \right) \right)&=\underset{n\to \infty }{\mathop{\lim }}\,\frac{1}{n}\underset{\forall {{\rho }_{A}}{{\rho }_{B}}}{\mathop{\max }}\,{{I}_{coh}}\left( {{\rho }_{{{\mathcal{G}}_{E}}{{E}_{1}}}} \right) \\ 
 & =\underset{n\to \infty }{\mathop{\lim }}\,\frac{1}{n}\underset{\forall {{\rho }_{A}}{{\rho }_{B}}}{\mathop{\max }}\,\left( \mathcal{D}\left( {{\rho }_{{{\mathcal{G}}_{E}}{{E}_{1}}}} \right)+\mathcal{C}\left( {{\rho }_{{{\mathcal{G}}_{E}}{{E}_{1}}}} \right)-1 \right) \\ 
 & =\underset{n\to \infty }{\mathop{\lim }}\,\frac{1}{n}\underset{\forall {{\rho }_{{{\mathcal{G}}_{E}}{{E}_{1}}}}}{\mathop{\max }}\,\left( I\left( {{\rho }_{{{\mathcal{G}}_{E}}{{E}_{1}}}} \right)-1 \right) \\ 
 & =\underset{n\to \infty }{\mathop{\lim }}\,\frac{1}{n}\underset{\forall {{\rho }_{{{\mathcal{G}}_{E}}{{E}_{1}}}}}{\mathop{\max }}\,\left( \begin{split}
  & S\left( {{\rho }_{{{\mathcal{G}}_{E}}}} \right)+S\left( {{\rho }_{{{E}_{1}}}} \right)+{{u}_{+}}{{\log }_{2}}{{u}_{+}}+{{u}_{-}}{{\log }_{2}}{{u}_{-}}\\ 
 & +{{v}_{+}}{{\log }_{2}}{{v}_{+}}+{{v}_{-}}{{\log }_{2}}{{v}_{-}}-1 \\ 
\end{split} \right).  
\end{split}
\end{equation}
Because $\rho _{{\rm {\mathcal G}}_{E} E_{1} } $ is a Bell diagonal state with $r=s=0$, $S\left(\rho _{{\rm {\mathcal G}}_{E} } \right)=S\left(\rho _{E_{1} } \right)=1$, $Q\left({\rm {\mathcal M}}\left(B_{2} \right)\right)$ is simplified to
\begin{equation} \label{78)} 
Q\left({\rm {\mathcal M}}\left(B_{2} \right)\right)=\mathop{\lim }\limits_{n\to \infty } \frac{1}{n} \mathop{\max }\limits_{\forall \rho _{{\rm {\mathcal G}}_{E} E_{1} } } \left(1-S\left(\rho _{{\rm {\mathcal G}}_{E} E_{1} } \right)\right). 
\end{equation} 
The results of the correlation measure analysis are summarized in \fref{fig4}. 

 \begin{center}
\begin{figure*}[h!]
\begin{center}
\includegraphics[angle = 0,width=1\linewidth]{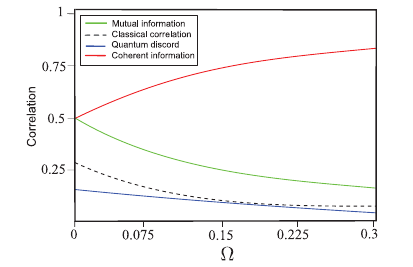}
\caption{The correlation measures between the quantum gravity environment ${\rm {\mathcal G}}_{E} $ and the local environment $E_{1} $, evaluated on $\rho _{{\rm {\mathcal G}}_{E} E_{1} } $, in function of $\Omega $, $\Omega \le {1}/{3}$. As $\Omega $ increases, the quantum influences become stronger, and the coherent information strongly increases. (The coherent information is shown in the absolute value.)} 
 \label{fig4}
 \end{center}
\end{figure*}
\end{center}

The quantum gravity environment allows the transfer of classical and quantum information through the entangled partition ${\rm {\mathcal G}}_{E} -E_{i} B_{j} $ of $\rho _{{\rm {\mathcal G}}_{E} E_{i} B_{j} } $, which concludes that the local maps ${\rm {\mathcal M}}_{1} $ and ${\rm {\mathcal M}}_{2} $ can extract classical and quantum information from the quantum gravity environment. 
\end{proof}

\section{Stimulated Storage in Quantum Gravity Memories }
\label{sec4}
The quantum gravity scenario allows us to build quantum memories with a non-fixed causality. In this section, we propose an example for this statement. Our quantum gravity memory is a quantum SR latch (S---set, R---reset), built from a pair of cross-coupled Toffoli-NOR quantum gates.

In classical computer architectures, the SR latch (flip-flop or bistable multivibrator) is one of the most basic and fundamental storage elements and building blocks of digital electronics devices. An SR latch consists of two cross-coupled NOR gates for the storing of one-bit information, and it operates with two stable states. The SR latch has two control inputs and two signal inputs, which are the back-looped outputs of the neighboring NOR gate (called cross coupling). The output of the classical SR latch is controlled by the $S$ and $R$ inputs, which allows only one stable output realization, $Q$, or its complement, $\bar{Q}$. The state transitions of the cross-coupling structure have a fixed causal structure in a classical SR latch. 

In particular, in a quantum gravity SR latch, both output realizations are simultaneously allowed as stable state, which makes possible the stimulated storage of a qubit entanglement ${\left| \varphi  \right\rangle} ={\textstyle\frac{1}{\sqrt{2} }} \left({\left| Q\bar{Q} \right\rangle} +{\left| Q\bar{Q} \right\rangle} \right)$, utilizing the elements of the standard basis ${\left| A_{i}  \right\rangle} \in \left\{{\left| 0 \right\rangle} ,{\left| 1 \right\rangle} \right\}$ as inputs. The proposed quantum gravity SR latch exploits the information resource-pool property (see Theorem 2) of the quantum gravity space to preserve the entanglement. 

The $C_{Toff}^{NOR} $ Toffoli-NOR qubit gate with control qubit inputs $x$ and $y$ and a target qubit $z$ can be defined as
\begin{equation} \label{79)} 
C_{Toff}^{NOR} =G\left(x,y,z\right)=z\oplus \left(\bar{x}\cdot \bar{y}\right), 
\end{equation} 
where $G\left(\cdot \right)$ refers to gate, and $\oplus $ stands for the XOR-operation.

The $C_{Toff}^{NOR} $ quantum circuit can be characterized by the density
\begin{equation} \label{80)}
\begin{split}
   C_{Toff}^{NOR}=&\left| 000 \right\rangle \left\langle  001 \right|+\left| 001 \right\rangle \left\langle  000 \right|+\left| 010 \right\rangle \left\langle  010 \right|+\left| 011 \right\rangle \left\langle  011 \right|\\ 
  &+\left| 100 \right\rangle \left\langle  100 \right|+\left| 101 \right\rangle \left\langle  101 \right|+\left| 110 \right\rangle \left\langle  110 \right|+\left| 111 \right\rangle \left\langle  111 \right|.  
\end{split}
\end{equation}

\begin{center}
\begin{figure*}[h!]
\begin{center}
\includegraphics[angle = 0,width=1\linewidth]{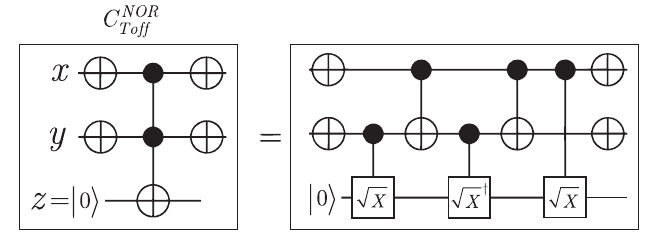}
\caption{The Toffoli-NOR qubit gate. The gate has two control qubit inputs $x$ and $y$ and a target qubit $z$, which is initialized in ${\left| 0 \right\rangle} $.} 
 \label{fig5}
 \end{center}
\end{figure*}
\end{center}

The $C_{Toff}^{NOR} $ structure can be decomposed into $NOT:a\to \bar{a}$, $CNOT:\left(a,b\right)\to \left(a,a\oplus b\right)$, and $\sqrt{X} $ and $\sqrt{X} ^{\dag } $ transformations, where 
\begin{equation} \label{82)}
\sqrt{X}=\frac{1}{2}\left( \begin{matrix}
   1+i & 1-i  \\
   1-i & 1+i  \\
\end{matrix} \right),
\end{equation}
and
\begin{equation} \label{83)}
{{\sqrt{X}}^{\dagger }}=\frac{1}{2}\left( \begin{matrix}
   1-i & 1+i  \\
   1+i & 1-i  \\
\end{matrix} \right).
\end{equation}
The $C_{Toff}^{NOR} $ Toffoli-NOR quantum circuit is shown in \fref{fig5}.

The truth table of the $C_{Toff}^{NOR} $ gate is given in \tref{tr1}.
 \begin{center}
\begin{table*}
\centering
\caption{The truth table of the Toffoli-NOR qubit gate.}
\label{tr1}
\begin{center}
\begin{tabular}
{|C{0.1in}|C{0.2in}|C{0.2in}|C{0.3in}|C{0.6in}|} \hline 
$x$ & y & $z$ & $\bar{x}\cdot \bar{y}$ & $z\oplus \left(\bar{x}\cdot \bar{y}\right)$ \\ \hline 
0 & 0 & 0 & 1 & 1 \\ \hline 
0 & 1 & 0 & 0 & 0 \\ \hline 
1 & 0 & 0 & 0 & 0 \\ \hline 
1 & 1 & 0 & 0 & 0 \\ \hline 
\end{tabular}
\end{center}
\end{table*}
\end{center} 
The ${\rm {\mathcal M}}_{SR} $ quantum gravity SR latch memory consists of two cross-coupled $C_{Toff}^{NOR} $ circuits, referred by the local maps ${\rm {\mathcal M}}_{S} $ and ${\rm {\mathcal M}}_{R} $ and defined by the following map:
\begin{equation} \label{ZEqnNum274102} 
{\rm {\mathcal M}}_{SR} \left(\rho \right)=\sum _{i}A_{i}^{SR} \rho \left(A_{i}^{SR} \right)^{\dag }  ,                             
\end{equation} 
where $S\in \left\{0,1\right\}$, $R\in \left\{0,1\right\}$. The map of ${\rm {\mathcal M}}_{SR} $ describes the parallel realizations of the local maps ${\rm {\mathcal M}}_{S} $ and ${\rm {\mathcal M}}_{R} $.

The Kraus operator $A_{i}^{SR} $ of \eqref{ZEqnNum274102} is expressed as 
\begin{equation} \label{85)} 
A_{i}^{SR} ={\left| 0 \right\rangle} {\left\langle 0 \right|} \otimes A_{i}^{A_{1} Q} \otimes A_{j}^{A_{2} \bar{Q}} +{\left| 1 \right\rangle} {\left\langle 1 \right|} \otimes A_{j}^{A_{2} \bar{Q}} \otimes A_{i}^{A_{1} Q} ,              
\end{equation} 
where $A_{i} \in \left\{0,1\right\}$ is the local input, $Q\in \left\{0,1\right\}$ is the output ${\rm {\mathcal M}}_{R} $, $\bar{Q}\in \left\{0,1\right\}$ is the output ${\rm {\mathcal M}}_{S} $, and the Kraus operators of ${\rm {\mathcal M}}_{S} $ and ${\rm {\mathcal M}}_{R} $ are
\begin{equation} \label{86)} 
{\rm {\mathcal M}}_{S} \left(\rho \right)=\sum _{i}A_{i}^{A_{2} \bar{Q}}  \rho \left(A_{i}^{A_{2} \bar{Q}} \right)^{\dag } ,                            
\end{equation} 
\begin{equation} \label{87)} 
{\rm {\mathcal M}}_{R} \left(\rho \right)=\sum _{j}A_{j}^{A_{1} Q}  \rho \left(A_{j}^{A_{1} Q} \right)^{\dag } . 
\end{equation} 
The control inputs $S$, $R$ of ${\rm {\mathcal M}}_{S} $ and ${\rm {\mathcal M}}_{R} $ are entangled with the quantum gravity environment state ${\rm {\mathcal G}}_{E} $. In the quantum gravity SR latch, input $R$ is separable from ${\rm {\mathcal G}}_{E} S$ and input $S$ is separable from ${\rm {\mathcal G}}_{E} R$; however, ${\rm {\mathcal G}}_{E} $ is entangled with $SR$, formulating the tripartite system (see Theorem 1) 
\begin{equation} \label{ZEqnNum352168} 
\rho _{{\rm {\mathcal G}}_{E} RS} =\kappa \cdot \xi +\left(1-\kappa \right)\chi ,                            
\end{equation} 
where $\kappa \le {\textstyle\frac{1}{3}} $, following the structure of \eqref{ZEqnNum436494}.

The main contribution of the ${\rm {\mathcal M}}_{SR} $ quantum gravity SR latch is that the non-fixed causality of the ${\rm {\mathcal G}}_{E} $ quantum gravity structure leads to the simultaneous realizations of the $Q$ and $\bar{Q}$ outputs, which can be used as the stimulation and storage of qubit entanglement, utilizing the resource-pool property of quantum gravity (see Theorem 2).

The active SR control commands are as
\begin{equation} \label{89)} 
{\left| S \right\rangle} :{\left| Q \right\rangle} ={\left| 1 \right\rangle} ,{\left| \bar{Q} \right\rangle} ={\left| 0 \right\rangle}  
\end{equation} 
and 
\begin{equation} \label{90)} 
{\left| R \right\rangle} :{\left| Q \right\rangle} ={\left| 0 \right\rangle} ,{\left| \bar{Q} \right\rangle} ={\left| 1 \right\rangle} ,                           
\end{equation} 
and in terms of the control state formalism, the realizations of the local maps is $C={\left| 0 \right\rangle} :{\left| S\bar{R} \right\rangle} $ and $C={\left| 1 \right\rangle} :{\left| \bar{S}R \right\rangle} $.

The truth table of the ${\rm {\mathcal M}}_{SR} $ quantum-gravity SR-latch is given in \tref{tr2}.
 \begin{center}
\begin{table*}
\centering
\caption{The truth table of the quantum-gravity SR-latch.}
\label{tr2}
\begin{center}
\begin{tabular}
{|C{0.2in}|C{0.2in}|C{0.2in}|C{0.3in}|C{0.3in}|} \hline 
$C$ & $S$ & $R$ & $Q$ & $\bar{Q}$ \\ \hline 
0 & 1 & 0 & 1 & 0 \\ \hline 
1 & 0 & 1 & 0 & 1 \\ \hline 
\end{tabular}
\end{center}
\end{table*}
\end{center}

\begin{center}
\begin{figure*}[h!]
\begin{center}
\includegraphics[angle = 0,width=1\linewidth]{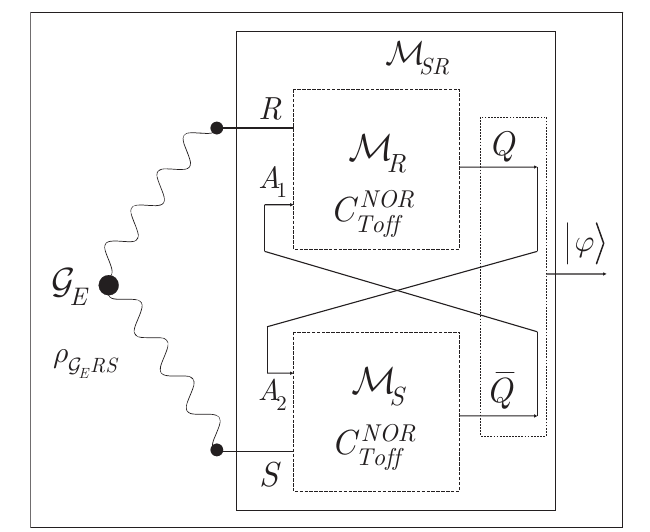}
\caption{Stimulated storage via the ${\rm {\mathcal M}}_{SR} $ quantum gravity SR latch memory (S---set, R---reset). The $S$ and $R$ inputs are controlled by the quantum gravity environment, formulating the tripartite system $\rho _{{\rm {\mathcal G}}_{E} RS} $ with entangled partition ${\rm {\mathcal G}}_{E} -RS$. The non-fixed causality of the quantum gravity structure leads to the parallel realizations of maps ${\rm {\mathcal M}}_{S} $ and ${\rm {\mathcal M}}_{R} $ (Toffoli-NOR gates) and the entanglement of $Q$ and $\bar{Q}$. The resource for the stimulation and storage processes is provided by the quantum gravity environment.} 
 \label{fig6}
 \end{center}
\end{figure*}
\end{center}

Initializing the circuit in ${\left| A_{1}  \right\rangle} ={\left| 0 \right\rangle} ,{\left| A_{2}  \right\rangle} ={\left| 0 \right\rangle} $ and by the control state (see \eqref{ZEqnNum707760}) ${\left| C \right\rangle} ={\textstyle\frac{1}{\sqrt{2} }} \left({\left| 0 \right\rangle} +{\left| 1 \right\rangle} \right)$, one obtains 
\begin{equation} \label{91)} 
{\left| C \right\rangle} ={\textstyle\frac{1}{\sqrt{2} }} \left({\left| S\bar{R} \right\rangle} +{\left| \bar{S}R \right\rangle} \right),                             
\end{equation} 
thus, the resulting output of ${\rm {\mathcal M}}_{SR} $ is evaluated as
\begin{equation} \label{ZEqnNum675084} 
{\left| \varphi  \right\rangle} ={\textstyle\frac{1}{\sqrt{2} }} \left({\left| Q\bar{Q} \right\rangle} +{\left| \bar{Q}Q \right\rangle} \right).                             
\end{equation} 
The ${\rm {\mathcal M}}_{SR} $ quantum gravity SR-latch with quantum gravity control is depicted in \fref{fig6}. The system is initialized with inputs $A_{i} \in \left\{0,1\right\}$. The outputs $Q$ and $\bar{Q}$ are entangled, stimulated, and kept in a stable state by the quantum gravity space $\rho _{{\rm {\mathcal G}}_{E} RS} $. 

In this section, we showed that the information resource-pool property of the quantum gravity environment can be exploited in quantum memories. We proposed a quantum gravity memory device and introduced the term stimulated storage, which allows the stimulation and storage of qubit entanglement, exploiting the information resource-pool property of the quantum gravity environment. 

The results indicate that the structure of the quantum gravity space can be further exploited in the development of quantum devices and quantum computers.

\section{Conclusions}
\label{sec5}
The theory of quantum gravity integrates the fundamental results of quantum mechanics with general relativity. This fusion injects and adds several benefits to quantum mechanics, most importantly the non-fixed causality structure of space-time geometry and the existence of causally non-separable processes. Here, we provided a model for the information processing structure of the quantum gravity space. We analyzed the connection of the gravity environment with the local processes and revealed that the quantum gravity environment is an information transfer device. This property makes the use of quantum gravity space as an information resource-pool available for the parties. We introduced the term remote simulation and showed that the quantum gravity space induces noise on the local environment states, which allows the parties to simulate locally separated remote systems. We investigated the terms of quantum gravity memory and stimulated storage, which allows for the generation and preservation of the entanglement of qubits exploiting the information resource-pool property of quantum gravity. The information processing structure of quantum gravity can be further exploited in quantum computations, in quantum error correction, in quantum AI, in quantum devices, and particularly in the development of quantum computers.

\section*{Acknowledgements}
The research reported in this paper has been supported by the Hungarian Academy of Sciences (MTA Premium Postdoctoral Research Program 2019), by the National Research, Development and Innovation Fund (TUDFO/51757/2019-ITM, Thematic Excellence Program), by the National Research Development and Innovation Office of Hungary (Project No. 2017-1.2.1-NKP-2017-00001), by the Hungarian Scientific Research Fund - OTKA K-112125 and in part by the BME Artificial Intelligence FIKP grant of EMMI (Budapest University of Technology, BME FIKP-MI/SC).


\newpage
\appendix
\setcounter{table}{0}
\setcounter{figure}{0}
\setcounter{equation}{0}
\setcounter{algocf}{0}
\renewcommand{\thetable}{\Alph{section}.\arabic{table}}
\renewcommand{\thefigure}{\Alph{section}.\arabic{figure}}
\renewcommand{\theequation}{\Alph{section}.\arabic{equation}}
\renewcommand{\thealgocf}{\Alph{section}.\arabic{algocf}}

\setlength{\arrayrulewidth}{0.1mm}
\setlength{\tabcolsep}{5pt}
\renewcommand{\arraystretch}{1.5}
\section{Appendix}
\subsection{Abbreviations}
\begin{description}
\item[CNOT] Controlled-NOT
\item[CPTP] Completely Positive Trace Preserving
\item[GHZ] Greenberger–Horne–Zeilinger
\item[NOR] Negation of OR
\item[POVM] Positive Operator Valued Measure
\item[SR] Set-Reset
\end{description}

\subsection{Notations}
The notations of the manuscript are summarized in  \tref{tab2}.
\begin{center}
\begin{longtable}{||l|p{4.3in}||}
\caption{Summary of notations.}
\label{tab2}
\endfirsthead
\endhead
\hline
\textit{Notation} & \textit{Description} \\ \hline
${\rm {\mathcal M}}_{1} $,${\rm {\mathcal M}}_{2} $ & Independent local CPTP maps in the quantum gravity space. \\ \hline 
${\rm {\mathcal G}}_{E} $ & Quantum gravity environment (models the space-time geometry). \\ \hline 
$C\in \left\{{\left| 0 \right\rangle} ,{\left| 1 \right\rangle} \right\}$ & Controller state in a fixed causality. Controls the realization sequence of local maps. \\ \hline 
$C\in \left\{{\left| + \right\rangle} \right\}$ & Controller state in a non-fixed causality structure, ${\left| + \right\rangle} ={\textstyle\frac{1}{\sqrt{2} }} \left({\left| 0 \right\rangle} +{\left| 1 \right\rangle} \right)$. Models the vanishing causality between the local maps ${\rm {\mathcal M}}_{1} $, ${\rm {\mathcal M}}_{2} $ in the quantum gravity space. \\ \hline 
$B_{i} ,E_{i} $ & Local output and local environment state of a local CPTP map ${\rm {\mathcal M}}_{i} $. \\ \hline 
$B_{j} ,E_{j} $ & Remote output and environment state of a remote CPTP map ${\rm {\mathcal M}}_{j} $. \\ \hline 
$\rho _{{\rm {\mathcal G}}_{E} E_{i} B_{j} } $ & Entangled tripartite qubit system. Defines the entanglement structure of the space-time geometry with local environment $E_{i} $ and remote output $B_{j} $. \\ \hline 
$\rho $ & Density of parallel realizations of local maps ${\rm {\mathcal M}}_{1} $, ${\rm {\mathcal M}}_{2} $ in a non-fixed causality, $\rho ={\textstyle\frac{1}{2}} \rho _{{\rm {\mathcal G}}_{E} E_{1} B_{2} } +{\textstyle\frac{1}{2}} \rho _{{\rm {\mathcal G}}_{E} E_{2} B_{1} } $. \\ \hline 
$\left(\rho _{{\rm {\mathcal G}}_{E} E_{1} B_{2} } \right)^{T_{B_{2} } } $ & Partial transpose of $\rho _{{\rm {\mathcal G}}_{E} E_{1} B_{2} } $, with respect to subsystem $B_{2} $. If $\left(\rho _{{\rm {\mathcal G}}_{E} E_{1} B_{2} } \right)^{T_{B_{2} } } \ge 0$, then $B_{2} $ is separable from ${\rm {\mathcal G}}_{E} E_{1} $, while for $\left(\rho _{{\rm {\mathcal G}}_{E} E_{1} B_{2} } \right)^{T_{B_{2} } } <0$, the partition $B_{2} -{\rm {\mathcal G}}_{E} E_{1} $ is entangled. \\ \hline 
${{\mathcal{S}}_{P}}$ & Partition set of $\rho _{{\rm {\mathcal G}}_{E} E_{i} B_{j} } $, $ {{\mathcal{S}}_{P}}= \{ {{\rm {\mathcal G}}_{E} -E_{i} B_{j}, E_{i} -{\rm {\mathcal G}}_{E} B_{j}, B_{j} -{\rm {\mathcal G}}_{E} E_{i}} \} $.
Partition ${\rm {\mathcal G}}_{E} -E_{i} B_{j} $ is entangled, partitions $E_{i} -{\rm {\mathcal G}}_{E} B_{j} $, $B_{j} -{\rm {\mathcal G}}_{E} E_{i} $ are separable. Partition ${\rm {\mathcal G}}_{E} -E_{i} B_{j} $ models the entangled space-time geometry of the quantum gravity space. \\ \hline 
${\rm {\mathcal N}}_{A_{1} B_{1} } $, ${\rm {\mathcal N}}_{A_{2} B_{2} } $ & Local logical channels of maps ${\rm {\mathcal M}}_{1} $ and ${\rm {\mathcal M}}_{2} $, defined by Kraus operators ${\rm {\mathcal N}}_{A_{1} B_{1} } \left(\rho \right)=\sum _{j}A_{j}^{A_{1} B_{1} }  \rho \left(A_{j}^{A_{1} B_{1} } \right)^{\dag } $ and ${\rm {\mathcal N}}_{A_{2} B_{2} } \left(\rho \right)=\sum _{j}A_{j}^{A_{2} B_{2} }  \rho \left(A_{j}^{A_{2} B_{2} } \right)^{\dag } $. \\ \hline 
${\rm {\mathcal N}}_{A_{1} E_{1} } $, ${\rm {\mathcal N}}_{A_{2} E_{2} } $ & Local complementary channels of maps ${\rm {\mathcal M}}_{1} $,${\rm {\mathcal M}}_{2} $, defined via Kraus operators ${\rm {\mathcal N}}_{A_{1} E_{1} } \left(\rho \right)=\sum _{j}A_{j}^{A_{1} E_{1} }  \rho \left(A_{j}^{A_{1} E_{1} } \right)^{\dag } $ and ${\rm {\mathcal N}}_{A_{2} E_{2} } \left(\rho \right)=\sum _{j}A_{j}^{A_{2} E_{2} }  \rho \left(A_{j}^{A_{2} E_{2} } \right)^{\dag } $. \\ \hline 
${\rm {\mathcal D}}^{E_{i} \to B_{j} } $ & Local simulation map. Allows the remote simulation of remote output $B_{j} $ from the local environment state $E_{i} $, as $B_{j} =E_{i} \circ {\rm {\mathcal D}}^{E_{i} \to B_{j} } $ through the quantum gravity environment. The existence of ${\rm {\mathcal D}}^{E_{i} \to B_{j} } $ is the consequence of the entangled space-time geometry $\rho _{{\rm {\mathcal G}}_{E} E_{1} B_{2} } $. \\ \hline 
${\rm {\mathcal M}}_{{\rm {\mathcal G}}} $ & CPTP map which models the simultaneous realizations of the local channels ${\rm {\mathcal N}}_{A_{1} E_{1} } $, ${\rm {\mathcal N}}_{A_{2} B_{2} } $, defined as ${\rm {\mathcal M}}_{{\rm {\mathcal G}}} \left(\rho \right)=\sum _{i,j}A_{i}^{{\rm {\mathcal G}}} \rho \left(A_{i}^{{\rm {\mathcal G}}} \right)^{\dag }  $. \\ \hline 
${\rm {\mathcal M}}_{{\rm {\mathcal D}}} $ & Local CPTP map, describes the probabilistic simulation via ${\rm {\mathcal D}}^{E_{i} \to B_{j} } $ on the local environment $E_{i} $ as ${\rm {\mathcal M}}_{{\rm {\mathcal D}}} =p{\rm {\mathcal D}}^{E_{i} \to B_{j} } +\left(1-p\right)I$. The output of the map is $B'_{j} ={\textstyle\frac{1}{2}} B_{j} +{\textstyle\frac{1}{2}} E_{i} .$ \\ \hline 
$\Pi ^{X} $, $\Pi ^{Z} $ & Projective measurement in the $X$ and $Z$ basis. \\ \hline 
$W^{B_{1} E_{1} B_{2} E_{2} } $ & Process matrix, describes the causality relations of the local maps ${\rm {\mathcal M}}_{1} $ and ${\rm {\mathcal M}}_{2} $ of $\rho _{{\rm {\mathcal G}}_{E} E_{i} B_{j} } $ in the quantum gravity space. \\ \hline 
${\rm {\mathcal M}}_{A_{i} B_{j} } $ & The quantum gravity channel. It has a logical channel ${\rm {\mathcal N}}_{A_{i} B_{j} } $, that exists between the local input $A_{i} $ and the remote output $B_{j} $, and a local complementary channel ${\rm {\mathcal N}}_{A_{i} E_{i} } $, which exists between the local input $A_{i} $ and the local environment state $E_{i} $. The logical channel ${\rm {\mathcal N}}_{A_{i} B_{j} } $ is called the remote logical channel of ${\rm {\mathcal M}}_{A_{i} B_{j} } $, ${\rm {\mathcal N}}_{A_{i} B_{j} } ={\rm {\mathcal N}}_{A_{i} E_{i} } \circ {\rm {\mathcal D}}^{E_{i} \to B_{j} } $. The remote logical channel exits with probability p. \\ \hline 
${\rm {\mathcal M}}\left(B_{2} \right)$ & CPTP map $M_{2} \to M_{2} $, which gets as input the remote output $B_{2} $, and outputs $B'_{2} =E_{1} \circ {\rm {\mathcal M}}_{{\rm {\mathcal D}}} $, where ${\rm {\mathcal M}}_{{\rm {\mathcal D}}} =p{\rm {\mathcal D}}^{E_{i} \to B_{j} } +\left(1-p\right)I$. \\ \hline 
$\rho _{{\rm {\mathcal G}}_{E} E_{1} } $ & Bell diagonal state to quantify the correlations that is transmitted via the quantum gravity space ${\rm {\mathcal G}}_{E} $. \\ \hline 
$u_{+} ,u_{-} ,v_{+} ,v_{-} $ & Eigenvalues of $\rho _{{\rm {\mathcal G}}_{E} E_{1} } $, $\max \left\{v_{+} ,v_{-} ,u_{+} ,u_{-} \right\}\le {\textstyle\frac{1}{2}} $. \\ \hline 
$c_{1} ,c_{2} ,c_{3} $ & Parameters defined from the eigenvalues $v_{+} ,v_{-} $ as $c_{1} =\left(v_{+} -v_{-} \right),$$c_{2} =-\left(v_{+} -v_{-} \right)$ and $c_{3} =1-2\cdot \left(v_{+} -v_{-} \right)=1+2\cdot c_{2} $, $\left|c_{1} \right|+\left|c_{2} \right|+\left|c_{3} \right|\le 1$. \\ \hline 
$I\left(\cdot \right)$ & Mutual information function. \\ \hline 
${\rm {\mathcal C}}\left(\cdot \right)$ & Classical correlation function. \\ \hline 
${\rm {\mathcal D}}\left(\cdot \right)$ & Quantum discord. \\ \hline 
$I_{coh} \left(\cdot \right)$ & Coherent information. \\ \hline 
$C\left({\rm {\mathcal M}}\left(B_{2} \right)\right)$, $Q\left({\rm {\mathcal M}}\left(B_{2} \right)\right)$ & Classical and quantum capacity of channel ${\rm {\mathcal M}}\left(B_{2} \right)$. \\ \hline 
$C_{Toff}^{NOR} $ & Toffoli-NOR qubit gate, defined as $C_{Toff}^{NOR} =G\left(x,y,z\right)=z\oplus \overline{x+y},$ where $x$ and $y$ are the control qubit inputs, $z$ is the target qubit. \\ \hline 
$\sqrt{X} $, $\sqrt{X} ^{\dag } $ & Square-root $X$ operation and its adjoint. \\ \hline 
${\rm {\mathcal M}}_{SR} $ & Quantum gravity SR-latch memory. Consist of two cross-coupled $C_{Toff}^{NOR} $ circuits, referred by the local maps ${\rm {\mathcal M}}_{S} $ and ${\rm {\mathcal M}}_{R} $, ${\rm {\mathcal M}}_{SR} \left(\rho \right)=\sum _{i,j}A_{i}^{SR} \rho \left(A_{i}^{SR} \right)^{\dag }  $. \\ \hline 
${\left| \varphi  \right\rangle} ={\textstyle\frac{1}{\sqrt{2} }} \left({\left| Q\bar{Q} \right\rangle} +{\left| \bar{Q}Q \right\rangle} \right)$ & Entanglement of qubit of outputs $Q$ and $\bar{Q}$  in the quantum gravity ${\rm {\mathcal M}}_{SR} $. \\ \hline
\end{longtable}
\end{center}

\begin{thebibliography}{10}
\bibitem{ref2} Hardy, L. Probability Theories with Dynamic Causal Structure: A New Framework for Quantum Gravity. \textit{arXiv:grqc/0509120} (2005).

\bibitem{ref3} Hardy, L. Quantum Reality, Relativistic Causality, and Closing the Epistemic Circle, \textit{Essays in Honour of Abner Shimony}, (Myrvold, W. C. and Christian, J. eds.), Springer (2009).

\bibitem{ref4} Hardy, L. Quantum gravity computers: On the theory of computation with indefinite causal structure. \textit{arXiv:quant-ph/0701019v1} (2007).

\bibitem{ref5} Hardy, L. Towards quantum gravity: a framework for probabilistic theories with non-fixed causal structure, \textit{J. Phys. A: Math. Theor.} 40, 3081 (2007).

\bibitem{ref6} Lloyd, S. The computational universe: quantum gravity from quantum computation, \textit{quant-ph/0501135} (2005).

\bibitem{ref7} Lloyd, S. \textit{Programming the Universe: A Quantum Computer Scientist Takes On the Cosmos} (Alfred A. Knopf, New York, 2006).

\bibitem{ref8} Lloyd, S., Maccone, L., Garcia-Patron, R., Giovannetti, V., Shikano, Y., Piandola, S., Rozema, L. A., Darabi, A., Soudagar, Y., Shalm, L. K. and Steinberg, A. M. Closed timelike curves via post-selection: theory and experimental demonstration. \textit{Phys. Rev. Lett.} 106, 040403 (2011).

\bibitem{ref9} Barrett, J., Linden, N., Massar, S., Pironio, S., Popescu, S. and Roberts, D. Nonlocal correlations as an information-theoretic resource. \textit{Phys. Rev. A} 71, 022101 (2005).

\bibitem{ref10} Oreshkov, O., Costa, F. and Brukner, C. Quantum correlations with no causal order, \textit{Nature Communications} 3, 1092, doi:10.1038/ncomms2076.\textit{ arXiv:1105.4464v3} (2012).

\bibitem{ref11} Chiribella, G., D'Ariano, G. M., Perinotti, P. and Valiron, B. Quantum computations without definite causal structure, \textit{arXiv:0912.0195v4} (2013).

\bibitem{ref12} Pawlowski, M., Paterek, T., Kaszlikowski, D., Scarani, V., Winter, A. and  Zukowski, M. Information Causality as a Physical Principle, arXiv:0912.0195v4 \textit{Nature} 461, 1101-1104 (2009). 

\bibitem{add1} Procopio, L. M. et al. Experimental superposition of orders of quantum gates. \textit{Nature communications}, 6: 7913, (2015).

\bibitem{add2} Guerin, P. A., Feix, A., Araujo, M. and Brukner, C. Exponential communication complexity advantage from quantum superposition of the direction of communication. \textit{Physical Review Letters}, 117(10), (2016).

\bibitem{add3} Rubino, G. et al. Experimental verification of an indefinite causal order. \textit{Science Advances}, 3(3), e1602589 (2017).

\bibitem{add4} Long, G. L. The general quantum interference principle and duality computer. \textit{Commun. Theor. Phys.} 45 (5), 825-844 (2006).

\bibitem{add5} Milz, S. et al. Entanglement, non-Markovianity, and causal non-separability. \textit{New Journal of Physics}, 20(3):033033 (2018).

\bibitem{add6} Bang, J. et al. Quantifiable simulation of quantum computation beyond stochastic ensemble computation. \textit{Advanced Quantum Technologies}, 1(2): 1800037 (2018). 

\bibitem{add7} Mahmud, N. et al. Scaling Reconfigurable Emulation of Quantum Algorithms at High-Precision and High-Throughput. \textit{Quantum Engineering}, Vol 1, Issue 1, e19, (2019).

\bibitem{nagyp} Nagy, P. and Tasnadi, P. Projectile solutions on Minkowski diagram, \textit{Il Nuovo Cimento} vol. 33 C, No 3, Maggio-Giugno, DOI 10.1393/ncc/i2010-10638-5 (2010).

\bibitem{ref1} Gyongyosi, L., Imre, S. and Nguyen, H. V. A Survey on Quantum Channel Capacities, \textit{IEEE Communications Surveys and Tutorials}, DOI: 10.1109/COMST.2017.2786748 (2018).

\bibitem{ref13} Deutsch, D. Quantum computational networks, \textit{Proc. Roy. Soc. Lond. A }425, 73 (1989).

\bibitem{ref14} Bernstein, E. and Vazirani, U. Quantum Complexity Theory, \textit{SIAM J. of Computing} 26, (1997).

\bibitem{ref15} Paternostro, M., Vitali, D., Gigan, S., Kim, M. S., Brukner, C., Eisert, J. and Aspelmeyer, M. Creating and Probing Multipartite Macroscopic Entanglement with Light, \textit{Phys. Rev. Lett.} 99, 250401 (2007).

\bibitem{ref16} O’Connell, A. D., Hofheinz, M., Ansmann, M., Bialczak, R. C.,  Lenander, M., Lucero, E., Neeley, M., Sank, D., Wang, H., Weides, M., Wenner, J., Martinis, J. M. and Cleland, A. N. Quantum ground state and single-phonon control of a mechanical resonator. \textit{Nature} 464, 697 (2010).

\bibitem{ref17} Lee, K. C., Sprague, M. R., Sussman, B.J., Nunn, J., Langford, N. K., Jin, X. M., Champion, T., Michelberger, P., Reim, K. F., England, D., Jaksch, D. and Walmsley, I. A. Entangling macroscopic diamonds at room temperature. \textit{Science} 334, 1253 (2011).

\bibitem{ref18} Colnaghi, T., D'Ariano, G. M., Perinotti, P. and Facchini, S. Quantum computation with programmable connections between gates, \textit{Phys. Lett. A} 376, 2940 (2012).

\bibitem{ref19} Jencova, A. Generalized channels: channels for convex subsets of the state space, \textit{J. Math. Phys.} 53, 012201 (2012).

\bibitem{ref20} Gutoski, G. Properties of local quantum operations with shared entanglement. \textit{Quant. Inf. Comp.} 9, 739, (2009).

\bibitem{ref21} Aaronson, S. Quantum computing, postselection, and probabilistic polynomial-time, \textit{Proc. R. Soc. A} 461 3473 (2005).

\bibitem{ref22} Genkina, D., Chiribella, G. and Hardy, L. Optimal Probabilistic Simulation of Quantum Channels from the Future to the Past, \textit{Phys. Rev. A} 85, 022330 (2012).

\bibitem{ref23} Deutsch, D. Quantum mechanics near closed timelike lines. \textit{Phys. Rev. D} 44, 3197-3217 (1991).

\bibitem{ref24} Greenberger, D. M. and Svozil, K. Quantum Theory Looks at Time Travel. \textit{Quo Vadis Quantum Mechanics?}, Eds. Elitzur, A., Dolev, S. and Kolenda, N., Springer Verlag, Berlin (2005).

\bibitem{ref25} Navascues, M. and Wunderlich, H. A glance beyond the quantum model. \textit{Proc. Roy. Soc. Lond. A} 466, 881-890 (2009).

\bibitem{ref26} Wolf, M. M., Perez-Garcia, D. and Fernandez, C. Measurements Incompatible in Quantum Theory Cannot Be Measured Jointly in Any Other No-Signaling Theory. \textit{Phys. Rev. Lett.} 103, 230402 (2009).

\bibitem{ref27} Barnum, H. et al. Local Quantum Measurement and No-Signaling Imply Quantum Correlations. \textit{Phys. Rev. Lett.} 104, 140401 (2010).

\bibitem{ref27b} Acin, A. et al. Unified Framework for Correlations in Terms of Local Quantum Observables. \textit{Phys. Rev. Lett.} 104, 140404 (2010).

\bibitem{ref29} DeWitt, B. S. Quantum Theory of Gravity. I. The Canonical Theory. \textit{Phys. Rev.} 160, 1113-1148 (1967).

\bibitem{ref30} Peres, A. Measurement of time by quantum clocks. \textit{Am. J. Phys.} 48, 552-557 (1980).

\bibitem{ref31} Wooters, W. K. ``Time'' replaced by quantum correlations. \textit{Int. J. Theor. Phys.} 23, 701-711 (1984).

\bibitem{ref32} Isham, C. J. and Kuchar, K. V. Representations of Spacetime Dieomorphisms. 2. Canonical Geometrodynamics.\textit{ Ann. Phys. }164, 2, 316-333 (1985).

\bibitem{ref33} Gambini, R., Porto, R. A. and Pullin, J. A relational solution to the problem of time in quantum mechanics and quantum gravity: a fundamental mechanism for quantum decoherence. \textit{New J. Phys. }6, 45 (2004).

\bibitem{ref34} Gyongyosi, L. and Imre, S. A Survey on Quantum Computing Technology, \textit{Computer Science Review}, Elsevier, DOI: 10.1016/j. Cosrev.2018.11.002, ISSN: 1574-0137, (2018).

\bibitem{ref35} Imre, S. and Gyongyosi, L. \textit{Advanced Quantum Communications - An Engineering Approach}. Wiley-IEEE Press (New Jersey, USA), (2013).

\bibitem{ref36} Gyongyosi, L. The Correlation Conversion Property of Quantum Channels, \textit{Quantum Information Processing}, Springer, ISSN: 1570-0755 (print version), ISSN: 1573-1332 (2013).

\bibitem{ref37} Gyongyosi, L. and Imre, S. Distillable Entanglement from Classical Correlation, \textit{Proceedings of SPIE Quantum Information and Computation XI}, (2013).

\bibitem{ref38} Gyongyosi, L. The Structure and Quantum Capacity of a Partially Degradable Quantum Channel, \textit{IEEE Access}, ISSN: 2169-3536, \textit{arXiv:1304.5666 }(2014).

\bibitem{ref39} Gyongyosi, L. Quantum Information Transmission over a Partially Degradable Channel, \textit{IEEE Access}, ISSN: 2169-3536, (2014).

\bibitem{ref40} Gyongyosi, L. The Private Classical Capacity of a Partially Degradable Quantum Channel, \textit{Physica Scripta - Special Issue on Quantum Information}, Online ISSN: 1402-4896 Print ISSN: 0031-8949, (2014).

\bibitem{ref41} Choi, M. D. Completely positive linear maps on complex matrices, \textit{Linear Algebr. Appl.} 10, 285 (1975).

\bibitem{ref42} Cubitt, T. S., Ruskai, M. B. and Smith, G. The structure of degradable quantum channels, \textit{J. Math. Phys.} 49, 102104 (2008).

\bibitem{ref43} Oriti, D. \textit{Approaches to Quantum Gravity: Toward a New Understanding of Space, Time and Matter}, Cambridge Univ. Press, Cambridge, (2009).

\bibitem{ref44} Piazza, F. Glimmers of a Pre-geometric Perspective.\textit{ Found. Phys}. 40, 239-266 (2010).

\bibitem{ref45} Zurek, W. H. Decoherence and the transition from quantum to classical. \textit{Phys. Today} 44, 36-44 (1991).

\bibitem{ref46} Kofler, J. and Brukner, C. Classical world arising out of quantum physics under the restriction of coarse-grained measurements. \textit{Phys. Rev. Lett}. 99, 180403 (2007).

\bibitem{ref47} Bombelli, L., Lee, J. H., Meyer, D. and Sorkin, R. Space-time as a causal set.\textit{ Phys. Rev. Lett.} 59, 521-524 (1987).

\bibitem{ref48} D'Ariano, G. M. and Tosini, A. Space-time and special relativity from causal networks. \textit{arXiv:1008.4805} (2010).

\bibitem{ref49} Hawking, S. W., King, A. R. and McCarthy, P. J. A new topology for curved space-time which incorporates the causal, differential, and conformal structures. \textit{J. Math. Phys.} 17, 174-181 (1976).

\bibitem{ref50} Malament, D. B. The class of continuous timelike curves determines the topology of spacetime. \textit{J. Math. Phys.} 18, 1399-1404 (1977).

\bibitem{ref51} Bennett, C. H., Leung, D., Smith, G. and Smolin, J. Can Closed Timelike Curves or Nonlinear Quantum Mechanics Improve Quantum State Discrimination or Help Solve Hard Problems? \textit{Phys. Rev. Lett}. 103, 170502 (2009).

\bibitem{ref52} Huang, Y. Quantum discord for two-qubit X states: Analytical formula with very small worst-case error, \textit{Phys. Rev. A} 88, 014302 (2013).

\bibitem{ref53} Petz, D. \textit{Quantum Information Theory and Quantum Statistics}, Springer-Verlag, Heidelberg, Hiv: 6. (2008).

\bibitem{ref54} Gyongyosi, L. Smooth Entropy Transfer of Quantum Gravity Information Processing, \textit{arXiv:1403.6717} (2014).

\bibitem{ref55} Gyongyosi, L. Correlation Measure Equivalence in Dynamic Causal Structures, \textit{arXiv:1603.02416} (2016).

\bibitem{pres} Preskill, J. Quantum Computing in the NISQ era and beyond, \textit{Quantum} 2, 79 (2018).

\bibitem{har} Harrow, A. W. and Montanaro, A. Quantum Computational Supremacy, \textit{Nature}, vol 549, pages 203-209 (2017).

\bibitem{aar} Aaronson, S. and Chen, L. Complexity-theoretic foundations of quantum supremacy experiments. \textit{Proceedings of the 32nd Computational Complexity Conference}, CCC '17, pages 22:1-22:67, (2017).

\bibitem{far1} Farhi, E. and Neven, H. Classification with Quantum Neural Networks on Near Term Processors, \textit{arXiv:1802.06002v1} (2018).

\bibitem{far2} Farhi, E., Goldstone, J., Gutmann, S. and Neven, H. Quantum Algorithms for Fixed Qubit Architectures. \textit{arXiv:1703.06199v1} (2017).

\bibitem{qcomputer} Arute, F. et al. Quantum supremacy using a programmable superconducting processor, \textit{Nature}, Vol 574, DOI:10.1038/s41586-019-1666-5 (2019).

\bibitem{refibm} IBM. \textit{A new way of thinking: The IBM quantum experience}. URL: http://www.research.ibm.com/quantum. (2017).

\bibitem{o2} Lloyd, S., Shapiro, J. H., Wong, F. N. C., Kumar, P., Shahriar, S. M. and Yuen, H. P. Infrastructure for the quantum Internet. \textit{ACM SIGCOMM} \textit{Computer} \textit{Communication Review}, 34, 9--20 (2004).

\bibitem{o7} Pirandola, S., Laurenza, R., Ottaviani, C. and Banchi, L. Fundamental limits of repeaterless quantum communications, \textit{Nature Communications}, 15043, doi:10.1038/ncomms15043 (2017).

\bibitem{o9} Pirandola, S., Braunstein, S. L., Laurenza, R., Ottaviani, C., Cope, T. P. W., Spedalieri, G. and Banchi, L. Theory of channel simulation and bounds for private communication, \textit{Quantum Sci. Technol}. 3, 035009 (2018).

\bibitem{o9b} Pirandola, S. Capacities of repeater-assisted quantum communications, \textit{arXiv:1601.00966} (2016).

\bibitem{o10} Laurenza, R. and Pirandola, S. General bounds for sender-receiver capacities in multipoint quantum communications, \textit{Phys. Rev. A} 96, 032318 (2017).

\bibitem{o11} Pirandola, S. End-to-end capacities of a quantum communication network, \textit{Commun. Phys.} 2, 51 (2019).

\bibitem{o13} Biamonte, J. et al. Quantum Machine Learning. \textit{Nature}, 549, 195-202 (2017). 

\bibitem{o14} Lloyd, S., Mohseni, M. and Rebentrost, P. Quantum algorithms for supervised and unsupervised machine learning. \textit{arXiv:1307.0411} (2013).

\bibitem{o15} Lloyd, S., Mohseni, M. and Rebentrost, P. Quantum principal component analysis. \textit{Nature Physics}, 10, 631 (2014).

\bibitem{o16} Lloyd, S. Capacity of the noisy quantum channel. \textit{Physical Rev. A}, 55:1613--1622 (1997).

\bibitem{o17} Lloyd, S. The Universe as Quantum Computer, \textit{A Computable Universe: Understanding and exploring Nature as computation}, Zenil, H. ed., World Scientific, Singapore, \textit{arXiv:1312.4455v1} (2013).

\bibitem{o18} Shor, P. W. Scheme for reducing decoherence in quantum computer memory. \textit{Phys. Rev. A}, 52, R2493-R2496 (1995).

\bibitem{o1} Van Meter, R. \textit{Quantum Networking}. ISBN 1118648927, 9781118648926, John Wiley and Sons Ltd (2014).

\bibitem{o6} Van Meter, R. and Devitt, S. J. Local and Distributed Quantum Computation, \textit{IEEE Computer} 49(9), 31-42 (2016).

\end{thebibliography}
\end{document}